\documentclass[a4paper,fleqn,usenatbib]{mnras}

\usepackage{newtxtext,newtxmath}

\usepackage[T1]{fontenc}
\usepackage{ae,aecompl}

\usepackage{graphicx}	
\usepackage{amsmath}	
\usepackage{amssymb}	

\title[WDs and two--planet systems instabilities]{Dynamical evolution of two-planet systems and its connection with white dwarf atmospheric pollution}

\author[R. F. Maldonado et al.]{
R. F. Maldonado,$^{1}$, E. Villaver $^{2}$, A. J. Mustill $^{3}$, M. Ch\'avez $^{1}$, E. Bertone $^{1}$  \\
\thanks{E-mail: raulfms@inaoep.mx}
$^{1}$Instituto Nacional de Astrof\'isica, \'Optica y Electr\'onica, Luis Enrique Erro 1, Tonantzintla, Puebla, M\'exico\\
$^{2}$Departamento de F\'isica Te\'orica, Universidad Aut\'onoma de Madrid, E-28049, Madrid, Spain\\
$^{3}$Lund Observatory, Box 43, SE-22100 Lund, Sweden\\
}
\date{Accepted XXX. Received YYY; in original form ZZZ}

\pubyear{2020}

\begin{document}
\label{firstpage}
\pagerange{\pageref{firstpage}--\pageref{lastpage}}
\maketitle


\begin{abstract}
Asteroid material is detected in white dwarfs (WDs) as atmospheric pollution by metals, in the form of gas/dust discs, or in photometric transits. Within the current paradigm, minor bodies need to be scattered, most likely by planets, into highly eccentric orbits where the material gets disrupted by tidal forces and then accreted onto the star. This can occur through a planet--planet scattering process triggered by the stellar mass loss during the post main--sequence evolution of planetary systems. So far, studies of the $N$-body dynamics of this process have used artificial planetary system architectures built ad hoc. In this work, we attempt to go a step further and study the dynamical instability provided by more restrictive systems, that, at the same time allow us an exploration of a wider parameter space: the hundreds of multiple planetary systems found around main--sequence (MS) stars. We find that most of our simulated systems remain stable during the MS,  Red and Asymptotic Giant Branch and for several Gyr into the WD phases of the host star. Overall, only $\approx$  2.3\,$\%$ of the simulated systems lose a planet on the WD  as a result of dynamical instability. If the instabilities take place during the WD phase most of them result in planet ejections with just 5 planetary configurations ending as a collision of a planet with the WD. Finally  3.2\,$\%$ of the simulated systems experience some form of orbital scattering or orbit crossing that could contribute to the pollution at a sustained rate if planetesimals are present in the same system.

\end{abstract}


\section{Introduction}

In 1917, van Maanen discovered the first  white dwarf (WD) that showed Ca II absorption lines in its spectrum: van Maanen 2 \citep{van1917}. Since then, 25\,$\%$ to 50\,$\%$ of the WDs  have been found with metallic lines (typically Mg, Si, Fe and other rock--forming elements) in their ultraviolet--optical spectra \citep{koester2014,harrison2018,wilson2019}. It is not expected that metallic lines are present in the spectra of cool WDs. The gravitational settling time, on which these elements sink out of the observable atmosphere, (a few days to years for DA WDs\footnote{WDs are classified according to the absorption lines present in their spectra, as DA, which are hydrogen dominated and DB, helium dominated.} and between few to million years in non--DA WDs \citealt{fontaine1979,wyatt2014}) is orders of magnitude shorter than the cooling time for these WDs ($t_\mathrm{cool}>100$ Myr for WDs with $T_\mathrm{eff}<$20000 K, \citealt{schreiber2003}). Thus, it was proposed that those cool metal--polluted  WDs should accrete material from the exterior \citep{alcock86,farihi2009,farihi2010,koester2014}. 

In the past, the processes invoked to explain the atmospheric pollution of WDs came in two flavours: either accretion from interstellar matter, or accretion of asteroidal material.  In the interstellar matter scenario \citep{dupuis1992,dupuis1993}, as the WD travels through the gravitational potential of the Galaxy, it passes through denser cloud regions and accretes metallic elements into its atmosphere. Nowadays, the commonly--accepted mechanism  involves  the presence of rocky bodies  that originally orbit the WD, are then dynamically delivered to the proximity of the WD, and finally subject to tidal forces that destroy them, allowing the accretion of material into the WD atmosphere. Observational evidence that supports this mechanism  includes: i) the near--infrared excesses that imply the presence of a dust disc located at few solar radii from the WD surface e.g. \citep{kilic2007}; ii) the double--peaked lines arising from keplerian rotation of gaseous material \citep[vaporised rock;][]{gansicke2006,melis2012,wilson2014,guo2015} observed in a few WDs; and iii) the fact that the chemical abundances found in these WDs resemble rocky material with bulk Earth composition \citep{jura2014,xu2014,harrison2018}. All this evidence is further confirmed by the direct probe of the presence of asteroid material detected as variable transits around the star WD~1145+017, explained as the disintegration of planetesimals  that have reached the WD's Roche limit \citep{vanderburg2015}, in addition to the recently--discovered transit of ZTF J013906.17+524536.89 by \cite{vanderbosh2019}. The planetesimal found orbiting the WD SDSS J122859.93+104032.9 with a 123.4--minute period \citep{manser2019} and the putative evaporating planet proposed to explain the accretion onto the WD J091405.30+191412.25 \citep{gansicke2019} are further proofs of the existence of asteroid and planetary bodies close to WDs. 

WDs represent the final stages in the lives of stars with masses between 1 and 8 $\mathrm{\,M}_\odot$. Stellar evolution (the large increase in stellar radius as a giant) and tidal forces should prevent the  survival of primordial planetary material up to a few \,au  from the WD surface \citep[see e.g.][]{villaver2009,mustill2012,villaver2014}.  Thus, material then needs to be delivered close to the WD likely by the scattering of small bodies from larger orbits when planetary systems are destabilized following the evolution of the star.  Pioneering work in the problem was done by \cite{duncan1998}, who showed that the planets of the Solar System would remain stable when the Sun becomes a WD, and by \citet{debes2002}, who explored the dynamical evolution of two-- and three--planet systems of identical mass in circular orbits before and after adiabatic mass loss. Here, the fundamental idea is that after the star loses mass at the AGB tip (around 75\,\% for a star of initially 3 $\mathrm{\,M}_\odot$), the planet:star mass ratio increases by a factor of a few compared to the original main--sequence (MS) configuration. Therefore, asteroids, planets or whole systems that survived the star's MS lifetime can be destabilised once the star becomes a WD.

The work of \cite{debes2002} has recently been extended, looking at a broader range of system architectures. \citet{veras2013} and \citet{veras2013b} looked at the stability of two--planet systems to the range  of  stellar masses that from the MS evolve to the WD (1 -- 8 $\mathrm{\,M}_\odot$) and at Jupiter and Earth--mass planets with five eccentricity values (0, 0.1, 0.2, 0.3, 0.5).  
The interaction of one planet with a belt of particles was explored by \citet{bonsor2011}, who reached the  conclusion that one planet does not seem to be enough to deliver material efficiently into the WDs unless the presence of a very massive belt is invoked. \cite{frewen2014} found single eccentric small planets to be more efficient, suggesting a successful mechanism to explain pollution of WDs; the question then becomes one of the origin of the planetary eccentricity. \cite{mustill2014} studied systems of triple Jupiter--mass planets, finding that the percentage of instabilities in the WD phase was insufficient to explain the observed pollution in host stars with mass between 3 -- 8 $\mathrm{\,M}_\odot$. Furthermore, the planet mass seems to have an effect on the instability outcome, as \citet{veras2016b} showed (using Jupiter, Saturn, Uranus and Neptune masses and circular orbits) that giant planets usually eject each other while smaller planets preferentially collide with each other. Recently, \citet{mustill2018} carried out simulations of three--planet systems with unequal--mass planets (from Super--Earth to Super--Jupiter masses) and including test particles to mimic planetesimals; they too showed that the lower--mass planets in the simulations deliver material towards the WD more efficiently. Most important is the fact that with low--mass planets \cite{mustill2018} found both higher rates and a longer duration of delivery, in line with the broad range of cooling ages at which metal pollution is observed.  

Building on previous works on the stability of multiple planetary systems using dynamical simulations,  the objective of this paper is twofold.  On the one hand, we aim at constraining the parameter space that previous studies could not, given the humongous number of parameters available if one builds the problem ad hoc. So we study the evolution to the WD phase of scaled versions of the MS planetary systems that have been detected, instead of using artificial planetary systems built for the problem. On the other hand, we allow an exploration of a larger parameter space than previous works, since we have configurations of planets with different masses, orbits, multiple eccentricities and with different semimajor axis ratios. Exploring all the range of parameters is otherwise unfeasible if not restricted by the system set--up built from the observed configurations. 

In this work, we use the orbital parameters of the hundreds of  multiple planetary systems with well--determined parameters found around MS stars, and explore their dynamical evolution to the WD phase. 
This is the first time dynamical simulations restricted by the observed parameters have been done  to study the instabilities that could bring material to the surface of the WD, thus producing the observed pollution.  In this paper, we focus on the two--planet systems; systems with three and more planets will be analyzed in a future study. In \S2 we describe how we have built the planetary systems to study, in \S3 we explain the scaling up of the planet mass and radius and the simulation set--up,  and in \S4 and \S5 we present the results and discussion of this work. Finally, in \S6 we summarize our conclusions.

\section{Simulations}
\label{setup}

In this work we take a novel approach to setting up the orbits of the planets in the systems we simulate. When studying the planetary systems that might be responsible for WD pollution, we are hindered by our lack of knowledge about wide--orbit planets orbiting intermediate--mass stars that are distant enough ($>$ few \,au) to survive the evolution of their host. Previous works have therefore constructed artificial systems of equal--mass planets \citep{debes2002,veras2013,mustill2014}, used the Solar System as a template \citep{veras2016b}, or constructed systems artificially from a prescribed distribution of planet masses and orbital spacings \citep{mustill2018}. Here we take a different approach: we use the large population of known multi--planet systems on closer orbits around lower--mass stars as templates for wider--orbit systems around intermediate--mass stars, scaling them up to maintain their dynamical properties. We describe this process in Section~\ref{sec:scaling}. In so doing, we are not asserting that wide--orbit planets around WD progenitors \emph{must} look like the better--studied population of close--in multiple--planet systems. Rather, we are constructing an artificial population that is somewhat grounded in reality, rather than prescribing masses and orbital separations as done previously.

To solve the dynamics of the systems, we use the  \textsc{Mercury} package \citep{chambers1999} in its modified version 
\citep{veras2013b,mustill2018}, which takes into account the change of the stellar mass and radius along the different evolution phases. We used the RADAU integrator with a tolerance parameter of 10$^{-11}$ as implemented in \citet{mustill2018}. We consider as a planet ejected when it reaches orbits above 1$\times 10^6$ \,au from the central star, and planets can also be removed due to planet--planet collisions and when they reach the stellar radius. 

In order to build the architectures of the planetary systems  we shall evolve, we have selected all the two--planet systems from the NASA Exoplanet Archive\footnote{https://exoplanetarchive.ipac.caltech.edu/} and the Exoplanet Encyclopedia\footnote{http://exoplanet.eu/} with reported discovery until June 2018 although we have updated (as of January 2020) the orbital parameters of some of the systems  (see Section \ref{msexp}). 
In those catalogues we have 29 multiple planetary systems in which a single host star has evolved beyond the MS according to the luminosity class as it appears in the {\it SIMBAD} database, the Exoplanet catalogues or the discovery paper. We have excluded from the simulations 13 giant stars, four Horizontal Branch stars and two pulsars for the simulations as the required treatment will be different from the rest of the simulations presented.  Although found in the catalogues as evolved stars, we have included the 10 subgiants that after verifying that they have not ascended yet the Giant branch in the HR diagram. One Herbig Ae/Be star has also been excluded from our sample. When the planetary system is orbiting one of the components of a binary star (39 stellar binary systems were identified) we excluded from the list cataclysmic binaries and eclipsing binaries (14 systems).  We also excluded 11 systems in which gravitational effects from a wide binary companion may affect the evolution of the planet orbits \citep[{\it Kepler}--108, HD 142, HD 89744, HD 133131A, HD 65216, HD 190360, HD 20781, XO--2S, HD 11964, HD 87646, GJ 229][]{moutou2017,otto2017,leggett2002}. We have kept in the simulations the systems orbiting HD 164922, HD 177830, HD 187123, HD 217107 and HD 143761 since \citet{wittrock2017} reported in their Differential Speckle Survey that they do not have a low mass stellar companion. We have also kept {\it Kepler}--383, {\it Kepler}--397, {\it Kepler}--400, {\it Kepler}--411, {\it Kepler}--449, {\it Kepler}--487, K2--36, HD 169830 and HD 147873 in our sample since we did not find any evidence in the literature that those are  physically related double systems.

The final sample we build for our study consists of 373 stars with two planets each. For those we select from the observations the stellar and planet masses, and all the orbital and planet parameters available.  

To build the simulations we have chosen an initial mass of the star of 3 $\mathrm{\,M}_\odot$. The mass of the host star of the observed systems ranges between 0.164 -- 1.965 $\mathrm{\,M}_\odot$. Thus, we have to re--scale the observed systems to a 3 $\mathrm{\,M}_\odot$ MS mass and we have to kept them dynamically analogous in order to evolve them. The choice of a 3 $\mathrm{\,M}_\odot$ is motivated by two facts, one observational and one computational. Polluted WDs have shown to have a mean mass of $\sim$ 0.7 $\mathrm{\,M}_\odot$ \citep{koester2014} which corresponds to a progenitor of 3 $\mathrm{\,M}_\odot$ mass on the MS following any standard initial--final mass relation (e.g. \citealt{kalirai2008}). Computationally, running a large number of dynamical simulations using low mass stars evolving off the MS become unfeasible given the long time the star expends on each evolutionary step. A 3 $\mathrm{\,M}_\odot$ star lives 377 Myr in the MS phase and at 477 Myr enters the WD phase [times derived using the SSE code of \citealt{hurley2000} which considers the isotropic mass loss during the Red Giant Branch (RGB) and Asymptotic Giant Branch (AGB) phases]. Note in contrast that a 1 $\mathrm{\,M}_\odot$ star requires $\sim$10 Gyr to leave the MS. The relatively rapid evolution of a 3 $\mathrm{\,M}_\odot$ star allows us to do a huge number of simulations in reasonable computational times (measured on a PhD timescale).  We note that orbital integrations speed up considerably once the star loses mass, as the orbits expand and the central mass is lower. Therefore, it is much quicker to run a system around a $3\mathrm{\,M}_\odot$ star for 10\,Gyr than to run one around a $1$ or $2\mathrm{\,M}_\odot$ star.

 We cannot be certain that the observed planet distribution matches that of the scaled $3\mathrm{\,M}_\odot$ mass stars we have simulated. The population of planets around WD stars is completely unknown yet and no planet has so far been confidently detected orbiting a WD despite systematic searches in the infrared and by transit \citep{burleigh2002,hogan2009,steele2011,mullally2008,debes2011,faedi2011,fulton14,vansluijs2018}.  Planets around A stars cannot be detected using the same techniques  as planets around G stars in the MS and therefore the limited statistics available for comparison among the different samples is subject to strong selection biases. Planet searches beyond the MS can give us some insight into the planet frequency around stars more massive than the Sun using the fact that the RV searches can be attempted once the star leaves the MS and its rotational velocity decreases. Early claims of an increase of planet mass with the mass of the host \citep[i.e.][]{lovis2007,johnson2007} are not supported by the results of more recent surveys that convincingly argue that this result might originate from limited RV precision and additional noise introduced by stellar p-mode oscillations \citep[see e.g.][]{niedzielski2016}. 

On the other hand, planet formation scenarios that investigated the frequency of giant planet formation with stellar mass find that the probability that a given star has at least one gas giant increases linearly with stellar mass from $0.4$ to $3\mathrm{\,M}_\odot$ \citep{kennedy2008} but planet multiplicity cannot be extracted from these models.  One could attempt to look  at the radius distribution of debris disks in A type stars compared to G stars but the radial extent of debris disks with well-resolved observations does not show any obvious trend with the stellar spectral type \citep[see e.g.][]{hughes2018}. But note that even if differences were found in terms of the size of the debris disks and the spectral type it will be very hard to attribute them to  planet formation \citep[e.g.][]{mustill2009} given that the size could be related as well to the location of the ice line \citep{morales2011} or to time effects in the production/destruction of dust \citep{kennedy2010}.

\section{The simulated architecture: Dynamically scaling the observed sample}

\label{sec:scaling}

To conserve the Hill stability criterion (see Section~\ref{equa} below) in the simulations with the adopted  $3\mathrm{\,M}_\odot$ star, we multiply the mass of each planet by a scale factor defined as $f=3\mathrm{\,M}_\odot/M_*$, where $M_*$ is the observed mass of the host star in the system. 

\subsection{Planet Masses and Radii }

The detection method determines the availability in the literature of distinct physical parameters of the planetary system. For instance, if the planet has been discovered by the transit method and has no radial velocity (RV) measurement then the radius of the planet is at hand but not its mass. Conversely, for those systems detected via RV  but that are not transiting we have (minimum) masses and eccentricities, but not radii. Thus, in order to complete the parameter space needed for the simulations, we need to use a planet mass--radius relation.  We have explored different prescriptions available in the literature and finally we adopted the one by \citet{chen2017}.  In that paper a planet mass--radius relation is defined by a probabilistic model of power laws at different mass regimes: for Earth--like worlds the function goes as $R\sim M^{0.28}$ (where $R$ and $M$ are the radius and mass of the planetary body respectively), for 
Neptunian worlds $R\sim M^{0.59}$, for Jovian exoplanets $R\sim M^{-0.04}$ and for stellar bodies $R\sim M^{0.88}$. We have used the \textsc{Python} package \textsc{Forecaster}  by the same authors  and assume that the standard deviation in the input parameter for the mass and radius is 5\,$\%$ 
Since \textsc{Forecaster} tests a variety of initial masses and radii, we opted to use the median after 100 runs.

We note that in a few cases (for some of the massive brown dwarfs which are listed in the exoplanet catalogues) the scaled--up mass of the ``planet'' is $>0.08\mathrm{\,M}_\odot$, making them large enough to burn hydrogen and become stars. We opted to keep these systems in the simulations in order to homogeneously treat our input catalogue.

\subsection{Initial Orbits}

\label{equa}
For two--planet systems, there exists an analytical criterion for whether the orbits of planets may intersect and the planets collide. This is the \emph{Hill stability limit} \citep{gladman1993}; in the following, we have used the two--body 
approximation for the energy, as given by \citet[and see also \citealt{veras2013b}]{donnison2011}.

In multiple--planet systems, stability can be classified in two ways: Hill and Lagrange instability.
In Hill--unstable systems, planets are close enough to have orbit crossing and to collide with each other. In Lagrange--unstable 
systems,  at least one planet is lost of the system via collision with the star or ejection outwards from the system. The Hill stability limit gives an analytic constraint on the necessary conditions that planetary systems need to have so that their planets collide or cross their orbits; however, the Lagrange stability limit does not have any analytic formulation and it can only be found through dynamical simulations. We are interested in exploring Hill-- and Lagrange--unstable systems to explain the atmospheric metal pollution observed in WDs.

Since \textsc{Mercury} does not take into account the stellar tidal forces that may act directly on the planets, 
we must ensure that these forces are negligible during the MS, RGB and AGB phases. For this reason we place 
the innermost planet to a semimajor axis $a_0$ = 10 \,au from the star since the surviving tidal 
limit for Jovian and Terrestrial planets is beyond this distance for a 3 $\mathrm{\,M}_\odot$ star 
\citep{villaver2009,mustill2012}. Setting up this limit at 10 \,au also allows us an easier comparison with previous works. 

Because the Hill stability limit depends directly on the semimajor axis ratio between the planets, we must keep the ratios of the observed planetary system in our simulations. 
Thus, the second planet is placed at a distance of $(a_2/a_1)a_0$ (where $a_2$ and $a_1$ are the observed semimajor axis of the outer and innermost planets respectively). For those planets for which the semimajor axis information was not available in the catalogues, we calculated it using the planet periods and the stellar mass (i.e. HD~114386, K2--141, {\it Kepler}--462 and {\it Kepler}--88).

\begin{figure*}
\begin{center}
\begin{tabular}{cc}
\includegraphics[width=9.0cm, height=7.5cm]{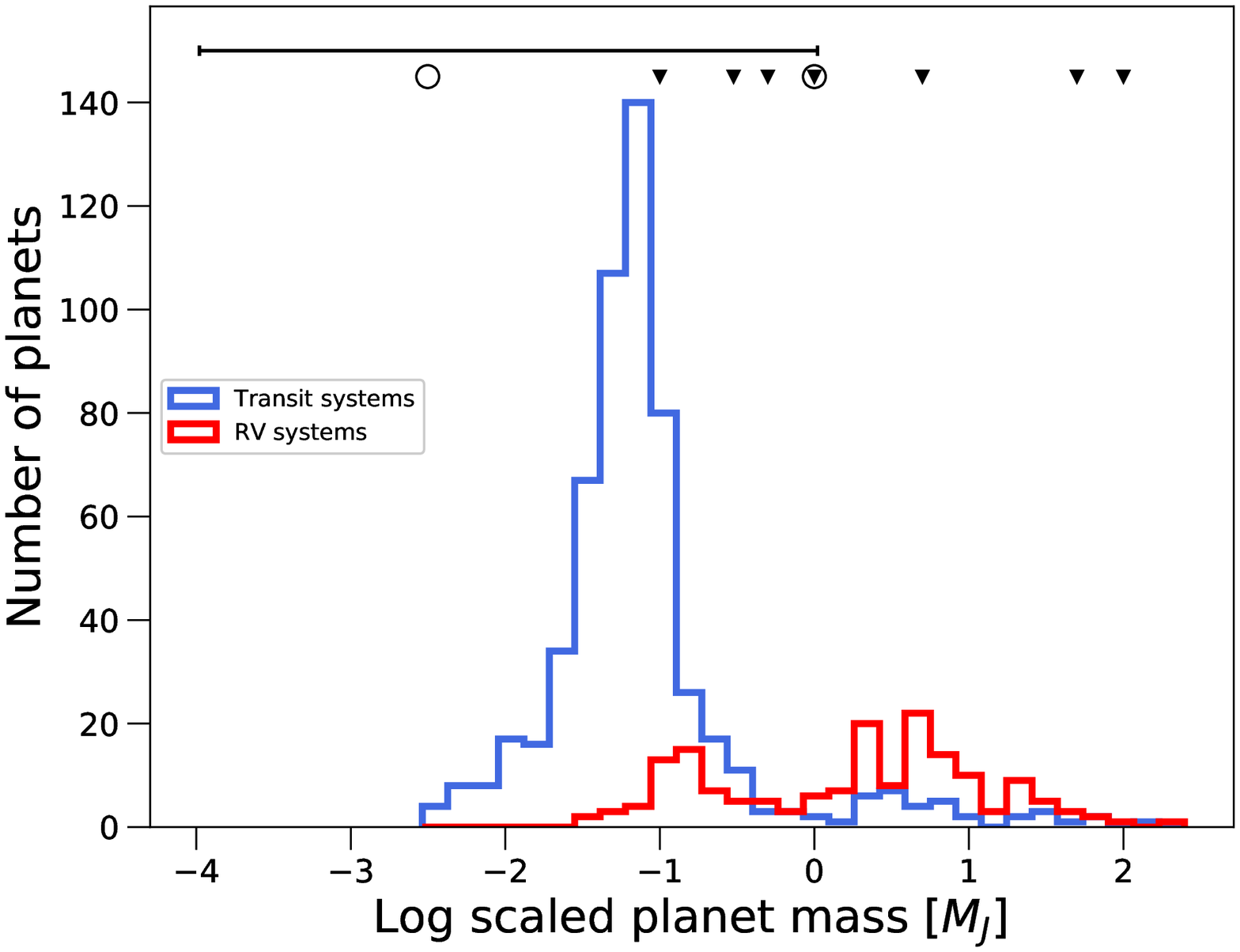}
\includegraphics[width=9.0cm, height=7.5cm]{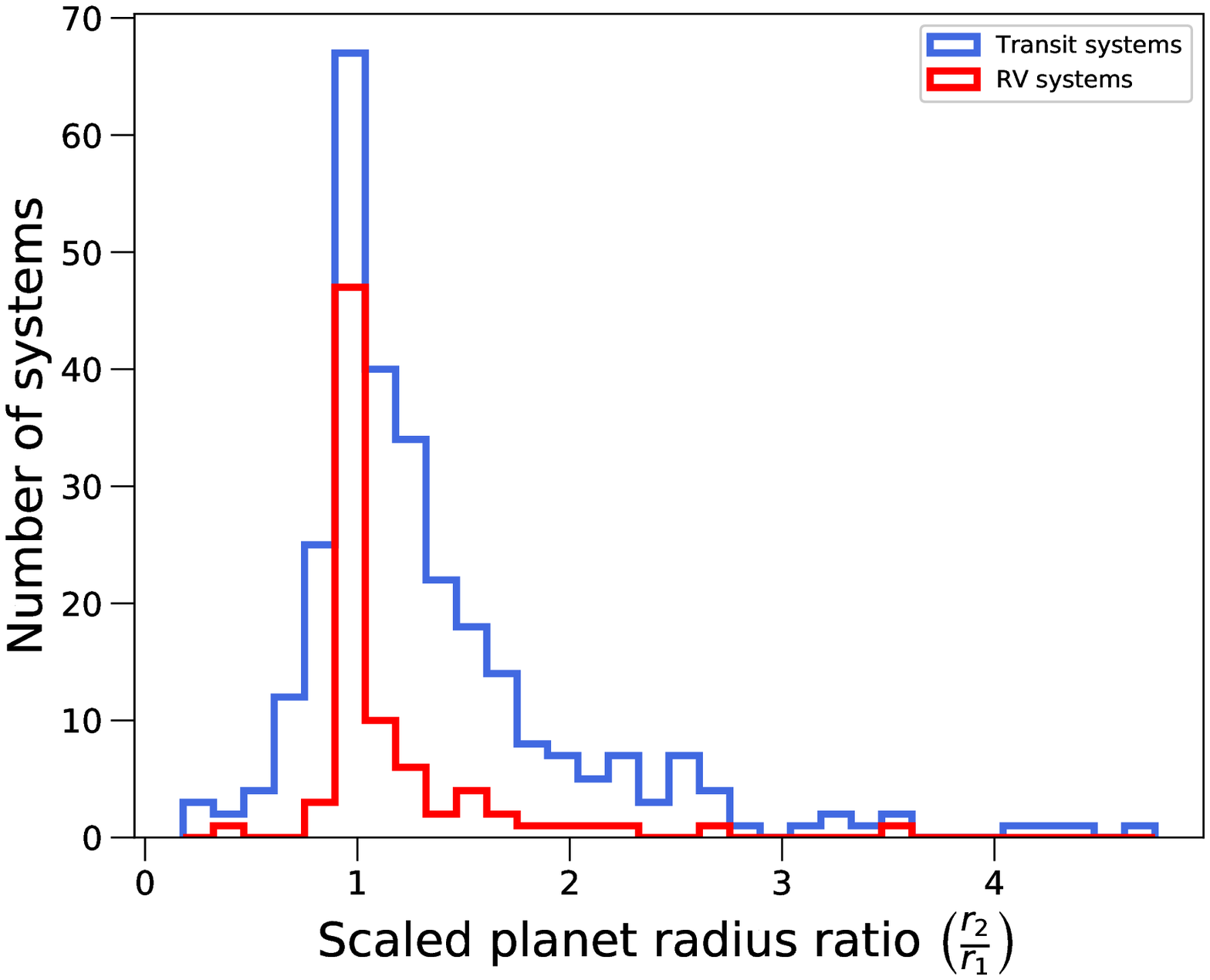} 
\end{tabular}
\caption{Scaled parameters of the planets considered in this work. Left, histogram of the scaled mass, showing 
in blue the Transit planets with calculated mass using the mass--radius relation by \citet{chen2017} and in red the RV systems with parameters reported in the catalogues. Black upside down triangles depict the masses used in \citet{veras2013} simulations, black circles the planet masses used in \citet{veras2013b} and in black horizontal line the planet mass range simulated in \citet{debes2002}.  The right panel shows  the 
distribution of the scaled planet radius ratio (outer/inner) of the Transit and RV systems. Same colours as the left panel.}
\label{4fig}
\end{center} 
\end{figure*} 

In Figure \ref{4fig} we present the distribution in mass (left) and radius ratio (right) of the planets we used in the simulations. In the left panel as the histogram of the scaled planet mass distribution  we see that the transiting systems show a single peak at around the mass of Neptune while the RV systems (which include some {\it Kepler} ones) have a bi--modal distribution with one peak close to the mass of Neptune and the other in the Jovian mass regime.  
In the top part of Fig. \ref{4fig} we have marked the values of the planet masses used in previous simulations where the triangles represent the planet masses used in \citet{veras2013}, circles are for \citet{veras2013b}, and the  horizontal line are the planet mass range covered by \citet{debes2002} in their two--planet simulations. This Figure clearly illustrates that we are exploring a new and extended parameter 
from previous works, specially for planets masses in between the ad hoc masses chosen in previous studies.

The right panel of Figure \ref{4fig} shows the distributions of the scaled radius ratio (outer/inner) of the systems we are simulating where a clear peak around 1 is present for both samples: an indication that planets in the same system have very similar sizes. Note that basically the same was found by \citet{weiss2018} in their analysis of the distribution of ratios of planet sizes for adjacent pairs within the same system observed by \emph{Kepler}. 

For the
calculation of the dynamics of the systems, \textsc{Mercury} requires in addition to the planet mass, radius and semimajor axis, the orbital eccentricity and 
inclination, the argument of the perihelion, mean anomaly and the longitude of the ascending node of 
each planet. The latter three angular parameters, since they are not available from the observations in most cases, are drawn randomly from a uniform distribution of angles between 0 and 360$^\circ$. The eccentricities are taken directly from the catalogues or, when unavailable, based on the results 
of \citet{vaneylen2015} and \cite{moorhead2011}, obtained from a  Rayleigh distribution 
with a $\sigma$ parameter $\sigma=0.02$ \citep{pu2015}. 
Regarding the inclination of the planet orbits, we have also randomly selected them from a Rayleigh distribution with $\sigma = 1.12^\circ$  \citep{xie2016}. 
The choice of using small inclination angles is justified in this work since  \citet{veras2018} concluded that near co--planar angles are adequate for global stability studies.

\begin{figure}
\begin{center}
\includegraphics[width=9cm, height=7.0cm]{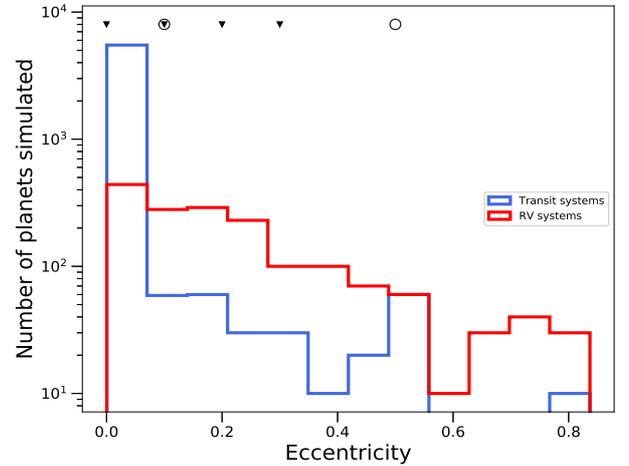} 
\caption{Histogram distribution of eccentricities of our simulated planets from transit detections (blue) and RV (red). The triangles and circles in the top of the panel are the eccentricities used in \citet{veras2013} and  \citet{veras2013b} respectively.}
\label{esemi}
\end{center} 
\end{figure}

In Figure \ref{esemi}, we show the histograms for the initial eccentricities of our two--planet simulated sample where the colours are as in previous Figures. Note that the numbers in this histogram are for simulated systems (as we explain later we perform 10 simulations per observed planetary system configuration). We see that the two samples peak at different eccentricities: the blue histogram simply reflecting the Rayleigh distribution used for those planets that did not have eccentricity measurements from observations (mostly transit systems) while the red histogram mimics the observed distribution from RV measurements. In the top of the Figure the triangles indicate the eccentricities used in \citet[$e=0,0.1, 0.2, 0.3$]{veras2013}, and the circles show the eccentricities used in the simulations of \citet[$e=0.1, 0.5$]{veras2013b}. Note that we have not simulated systems at zero eccentricity because this often reflects a lack of information, and it is more realistic to simulate them using a small eccentricity with a Rayleigh distribution with $\sigma =0.02$. Note as well that the our eccentricities cover a parameter space not studied in previous works. 

\begin{figure*}
\begin{center}
\includegraphics[width=14cm, height=12.5cm]{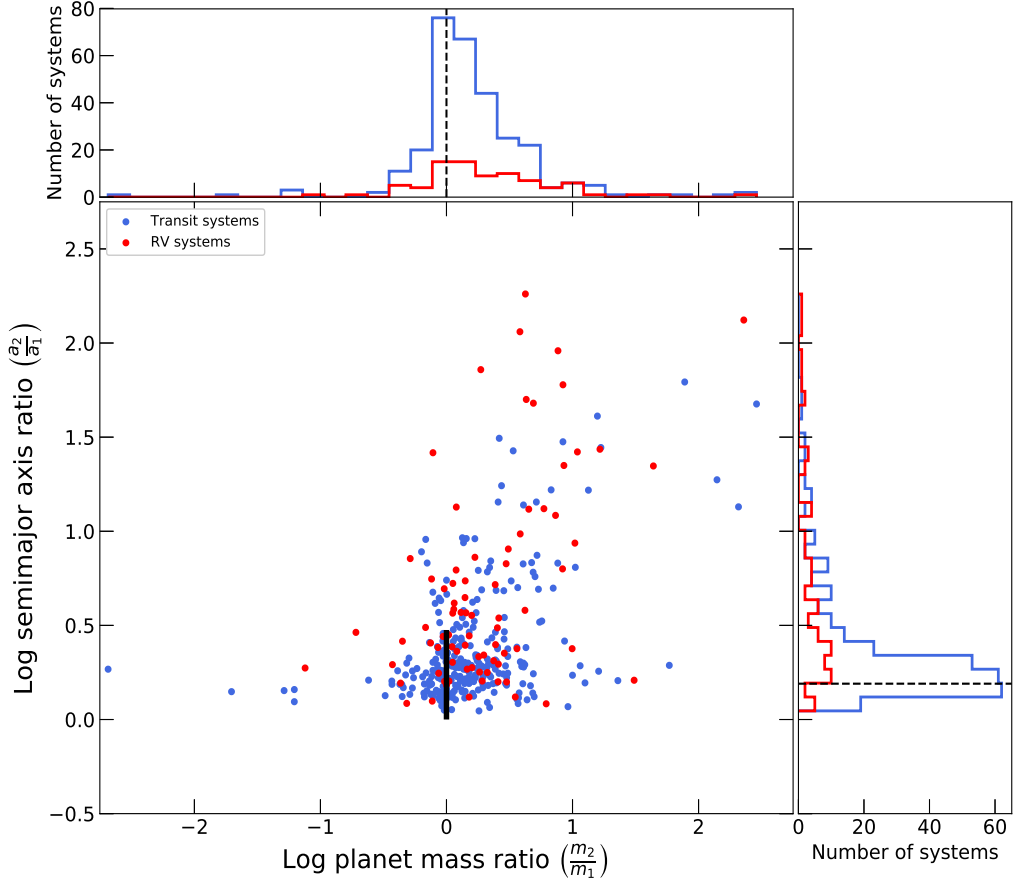} 
\caption{The center panel is a scatter plot of the semimajor axis ratio $a_2/a_1$ 
vs. the planet mass ratio $m_{2}/m_{1}$ of our simulated planetary systems. The black vertical line shows the equal mass planet ratio and its length represents the semimajor axis ratio range explored in \protect\cite{veras2013,veras2013b} for a 3$\mathrm{\,M}_\odot$ host star and for Jupiter--mass planets. In the top panel histogram we display the mass ratio distribution of our two--planet sample, where the black dashed line marks the location of two--planet systems with equal mass planets. The histogram in the right panel is the distribution of the semimajor axis ratio of our sample. The dashed vertical line corresponds to the semimajor axis ratio at which \protect\citet{veras2013b} found the Lagrange stability limit for their two--planet simulations, using equal mass Jupiter planets and eccentricity 0.1.}
\label{mrati}
\end{center} 
\end{figure*}

In Figure \ref{mrati} we display a scatter plot of the semimajor axis ratio versus the planet mass ratio. The black vertical line indicates the semimajor axis ratio range explored in \citet{veras2013,veras2013b} where they simulate two--planet systems with equal planet masses. In the upper part we show the histogram of the planet mass ratio distribution and in the right side of the scatter plot we have the histogram of the semimajor axis ratio. From this plot we see that both the parameter space in semimajor axis and planet mass ratio explored in this work is broader than previous works and that most systems have an outer planet (planet--2) of comparable mass but slightly more massive than the inner planet (planet--1).

\begin{figure}
\begin{center}
\begin{tabular}{c}
\includegraphics[width=9cm, height=7.5cm]{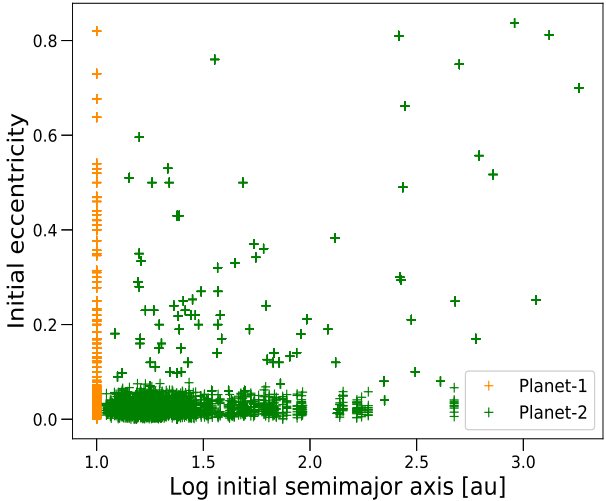} \\
\end{tabular}
\caption{Initial eccentricities as a function of initial semimajor axis of the two--planet systems simulated in this work. Orange and green plus symbols refers to planet--1 and --2 respectively. We note that both planets cover a wide range in eccentricities. Note that our planetary system configuration, locates planet--1 always at initial semimajor axis of 10 \,au while planet--2 will have the semimajor axis which correspond to the semimajor axis ratio of each observed system. }
\label{initea}
\end{center} 
\end{figure}

To finalize our description of the parameter set--up, we proceed to run 10 simulations per system configuration, changing randomly the inclination and eccentricity of the planet orbits using the Rayleigh distribution mentioned above and the orbital angles for each run. If the eccentricity is known, then we set it constant for the 10 simulations of the system. In Figure \ref{initea} we show the initial semimajor axis and eccentricity of the scaled two--planet systems simulated in this work. Orange and green plus symbols refer to planet--1 and planet--2 respectively. As we can see from our set--up, all planets--1 are located at 10 \,au, covering the eccentricity range from 0 to 0.8. On the other hand, planets--2 are widely dispersed and some of them are located at high eccentricities and large semimajor axis. We point out that these high eccentricity and high semimajor axis planets may come from systems that have likely already undergone an instability that we would think took place very early before the onset of our simulations.

\section{Results}

\begin{figure*}
\begin{center}
\begin{tabular}{cc}
\includegraphics[width=18cm, height=14cm]{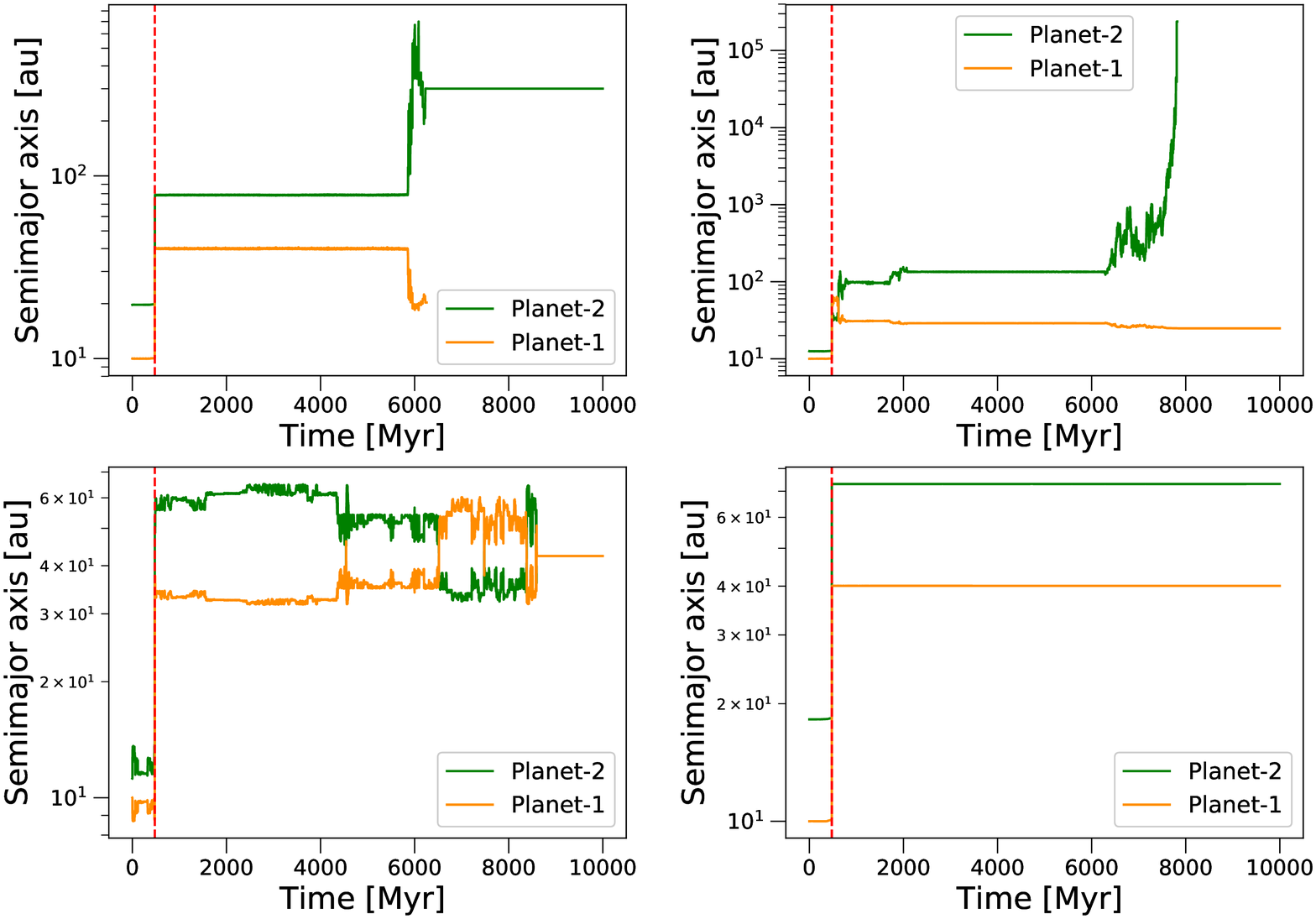} 
\end{tabular}
\caption{Evolution of the semimajor axis, in \,au, during the 10 Gyr of the simulations. In an  orange solid line we show the evolution of planet--1 while the  green solid line is for planet--2. The time when the star becomes a WD is shown as a  red vertical dashed line. The four panels are representative examples of the outcomes of the simulations  (\#293, 105, 1831, and 1465 left to right and top to bottom in Table \ref{mms}). Lagrange instability (planet--star collision) is shown in the top left panel,  Hill and Lagrange instabilities (orbit crossing and ejection of a planet) in the upper right panel. A Hill--unstable example (orbit crossing and collision between the planets) is the example shown in the bottom left panel and a complete Lagrange-- and Hill--stable system appear in the bottom right panel.} 
\label{semiaxis}
\end{center} 
\end{figure*}

We performed 3730 simulations of two--planet systems in which we
evolved a $3\mathrm{\,M}_\odot$ star from the MS to the WD. Before discussing the results in detail, we first show some examples of the types of dynamical evolution we see in our simulations. Such examples can be seen in Figure~\ref{semiaxis}.  Starting from the top left,  we have a system in which planet--1 is lost when it collides with the WD, in the top right a system that experiences orbit crossing and scattering followed by ejection of one planet, in the bottom left a system that underwent orbital crossing and a final planet--planet collision, and finally in the bottom right a fully stable system. 

The upper left panel of Figure \ref{semiaxis} shows one of the ten simulations of the scaled system HD~113538.  We see how planet--1 collides with the star at 6.2 Gyr, while planet--2 remains on a stable orbit at $a=300$ \,au after the instability.  Since this instability happens at a time when the star is well into the 
WD domain, this is a system in which we have a clear mechanism capable of producing pollution of the host star atmosphere, be it from the accreted planet itself or from asteroids scattered after the planet's eccentricities were excited. In the upper right panel of Figure \ref{semiaxis} we see the evolution of the scaled system GJ~180 showing  orbital scattering capable of producing pollution if the ejected planet--2 would traverse a planetesimal belt during scattering. In this system, the orbital scattering starts when the planets have an orbit crossing in the few Myr of the WD phase and the ejection of planet--2 happens at 7.8 Gyr.  The bottom left panel of Figure \ref{semiaxis} presents a collision between the planets in 
a simulation of the scaled system {\it Kepler}--200. The planet--planet collision happens at $t$= 8.6 Gyr, when the star has long been a WD, again capable of sending material into the WD atmosphere. This simulation also shows some orbit crossing of the planets, followed by scattering in their semimajor axis until the impact of the planets.  Note that in these examples, planets survive the MS phase and became unstable in the WD phase. Finally, the lower right panel evinces a fully stable evolution 
of the scaled system {\it Kepler}--146. In this case both planets are separated wide enough to be both Hill-- 
and Lagrange--stable during the complete simulated time of 10 Gyr, ending at 40 and 73 \,au respectively 
from the central star.  The input parameters used in the simulations shown in Figure \ref{semiaxis}  correspond to the simulation numbers 293, 105, 1831, and 1465 listed in the machine readable table, a fraction of which is displayed in Table \ref{mms}.

\begin{table*} 
\begin{center}
\small\addtolength{\tabcolsep}{-2pt}
\caption{Fraction of a machine readable table with the input parameters of the scaled systems of the 3730 simulations performed in this work.  Column 1: simulation number, 2: name of the planetary system, 3: scaled mass $m$, 4: planet density $\rho$, 5: scaled semimajor axis $a$, 6: eccentricity $e$, 7: orbit inclination $i$, 8: argument of the pericentre $\omega$, 9: longitude of the ascending node $\Omega$, 10: mean anomaly $M$.   The suffices 1 and 2 refer to the inner and outer planet respectively.  Column 11--18 the same but for planet--2. }
\label{mms}
\begin{tabular}{c c c c c c c c c c c c c c c c c c }
\noalign{\smallskip} \hline \noalign{\smallskip}
 $\#$ & name & $m_1$ & $\rho_1$  & $a_1$ & $e_1$ & $i_1$ & {$\omega_1$} &  $\Omega_1$  & $M_1$ & $m_2$ & $\rho_2$  & $a_2$ & $e_2$ & $i_2$ & {$\omega_2$} &  $\Omega_2$  & $M_2$   \\
 & & $[\mathrm{M_J}]$ & $[g/cm^3]$ & $[\,au]$ &  & $[^o]$ & $[^o]$ & $[^o]$ & $[^o]$ & $[\mathrm{M_J}]$ & $[g/cm^3]$ & $[\,au]$ &  & $[^o]$ & $[^o]$ & $[^o]$ & $[^o]$  \\
\noalign{\smallskip} \hline \noalign{\smallskip}
1	&	24Sex	&	3.88	&	3.17	&	10.0	&	0.09	&	0.29	&	14.42	&	184.86	&	287.96	&	1.68	&	1.22	&	15.60	&	0.29	&	1.24	&	217.37	&	353.30	&	302.54	\\
2	&	24Sex	&	3.88	&	3.17	&	10.0	&	0.09	&	0.57	&	188.29	&	37.46	&	204.71	&	1.68	&	1.22	&	15.60	&	0.29	&	1.43	&	357.32	&	318.77	&	317.99	\\
3	&	24Sex	&	3.88	&	3.17	&	10.0	&	0.09	&	0.79	&	302.46	&	200.24	&	138.43	&	1.68	&	1.22	&	15.60	&	0.29	&	1.45	&	241.91	&	236.67	&	69.66	\\
4	&	24Sex	&	3.88	&	3.17	&	10.0	&	0.09	&	1.25	&	45.52	&	52.84	&	264.22	&	1.68	&	1.22	&	15.60	&	0.29	&	3.07	&	76.19	&	254.02	&	51.88	\\
5	&	24Sex	&	3.88	&	3.17	&	10.0	&	0.09	&	1.37	&	120.95	&	252.23	&	297.85	&	1.68	&	1.22	&	15.60	&	0.29	&	1.36	&	196.07	&	151.61	&	65.64	\\
...	&	...	&	...	&	...	&	....	&	...	&	...	&	...	&	...	&	...	&	...	&	...	&	...	&	...	&	...	&	...	& ...	&	...	\\

\hline
\end{tabular}
\end{center}
\end{table*}

\subsection{Unstable Systems on the MS}
\label{msexp}

 A number of our two--planet systems (33) lost a planet on the MS before any stellar mass loss, frequently on a time--scale of just a few Myr. Because the template systems are observed at ages of typically a few Gyr, this means that the initial configurations for these systems were unphysical. We therefore investigated these more closely in order to identify any problems with the set--up.

23 of these systems lie close to strong first-- or second--order mean motion commensurabilities, most commonly the 2:1. In these cases a more careful set--up is required to place the system in a stable resonant configuration.  We find that 185 simulations out of these 23 systems, are having dynamical instabilities in the MS with 159 resulted in a planet being loss either by ejection, planet--planet or planet--star collisions, and 26 experienced orbit crossing. We defer the treatment of resonant two--planet systems experiencing stellar mass loss to a future work, but note that the stability properties of resonances are indeed known to change as the star loses mass and resonances broaden \citep{bonsor2011,mustillwyatt2012,debes2012,caiazzo2017}.

Besides the resonant systems, we found ten other systems experiencing instability on the MS:
\begin{itemize}
    \item \emph{HIP~57050:} This had a mass incorrectly listed as $68\mathrm{\,M_J}$ in the exoplanet.eu catalogue as of June 2018. This has since been corrected to $68\mathrm{\,M}_\oplus$ and we re--ran the simulations with the correct mass.
    \item \emph{HD~183263:} Planet--2 had an orbital solution proposed by \citet{wright2009}, and its  RV curve did not have a complete period; the solution is very close to being unstable. Neverthleess, \citet{feng2015} re--analyzed the RV curve updating the orbital parameters for planet--2 with a higher mass and longer period. We re-ran the simulations with the updated orbital solution.
    \item \emph{HD~202206:} This was listed as a two--planet system, but the innermost companion is in fact an M--dwarf. We have therefore removed it from consideration along with other binary stars.
    \item \emph{HD~67087:} Planet--2 has a high but poorly--constrained eccentricity ($0.76^{+0.17}_{-0.24}$) owing to the lack of observations at pericentre passage \citep{harakawa2015}. \cite{petrovich2015} identified it as unstable according to his stability criterion. Lacking a good orbital solution, we removed this system from further consideration.
    \item \emph{HD~106315:} \cite{crossfield2017} identified an RV trend indicating the presence of a third, outer, planet, and so we removed this system as it is probably not a two--planet system.
    \item \emph{HD~30177:} exoplanet.eu reported the unstable best fit from \cite{wittenmyer2017}; however; these authors identified a second, more stable, solution family with planet--2 on a wider orbit. We re--ran this system with the more stable configuration.
    \item \emph{Kepler--145:} Eccentricities are from photometry only and have large errors \citep{vaneylen2015}. We re--ran the system using instead the Rayleigh distribution of eccentricity which we used when the true eccentricity was unknown.
    \item \emph{Kepler--210:} \cite{ioannidis2014} provided a two--planet TTV fit which was unstable, and favoured instead a three--planet fit with the third planet not detected in transit. We therefore removed the system as it is probably not a two--planet system.
    \item \emph{K2--18:} exoplanet.eu reported high eccentricity values, but these were very poorly constrained by radial velocity measurements \citep{cloutier2017}. More recent work by \cite{sarkis2018} concluded that K2--18c was likely an artefact of stellar activity and therefore we removed this system from consideration.
    \item \emph{Kepler--462:} \cite{ahlers2015} provide a high lower limit ($>0.5$) on the eccentricity of planet--2 based on transit photometry, and noted that their solutions were unstable. Lacking a good fit, we remove this system from consideration.
\end{itemize}

\subsection{Global results}

We now consider a total of  3485 simulations for the following statistics since we have removed 6 two--planet systems (60 simulations) due to the previous analysis  and 185 additional simulations that experience dynamical instabilities on the MS. We keep the simulations of resonant systems that were stable on the MS, since their orbital configuration have lead them to avoid any destabilising effects of the resonances.

We find  101 (2.9\,$\%$) out of  3485 simulations that lose one planet in the simulations, be that by planet--planet collisions, planet--star collisions or ejections; while  3384 simulations (97.1\,$\%$) keep both planets the entire simulated time (10 Gyr). 

In Table \ref{tab} we show the numbers of planets lost, the type of dynamical instability by which it is removed from the system and two relative percentages, the first with respect to the total number of simulations and the second with respect to the total number of planets simulated (6970) at different evolutionary stages of the host star: MS phase ($t \leq 377.65$\,Myr); the pre--WD phase, which take into account the RGB and AGB phases ($377.65\mathrm{\,Myr} < t \leq 477.57$\,Myr) and the WD phase ($t > 477.57$\,Myr). 

\begin{table*} 
\begin{center}
\caption{Number of planet instabilities (collision between the planets, planet collision with the star, ejection) appearing at different evolutionary stages. The first percentage is the fraction of simulations in which the given outcome occurred; the second, the fraction of planets experiencing said outcome. ``Pre--WD'' means subgiant through to AGB tip.}
\label{tab}
\begin{tabular}{l c c c c}
\noalign{\smallskip} \hline \noalign{\smallskip}
 &  MS &  pre--WD  & WD  & Total \\
\noalign{\smallskip} \hline \noalign{\smallskip}
{\bf Ejections} & -- &  2  (0.06\,$\%$, 0.03\,$\%$) &  85 (2.44\,$\%$, 1.22\,$\%$) &  87 (2.5\,$\%$, 1.25\,$\%$) \\
{\bf Planet--star collisions} & -- & -- & 5 (0.14\,$\%$, 0.07\,$\%$) &  5 (0.14\,$\%$, 0.07\,$\%$)  \\
{\bf Planet--planet collisions} & -- &  2 (0.06\,$\%$, 0.03\,$\%$) &  7 (0.2\,$\%$, 0.1\,$\%$) &  9 (0.26\,$\%$, 0.13\,$\%$)\\
{\bf Total} & -- &  4 (0.12\,$\%$, 0.06\,$\%$) &  97 (2.78\,$\%$, 1.39\,$\%$) &  101 (2.9\,$\%$, 1.45\,$\%$) \\
\hline
\end{tabular}
\end{center}
\end{table*}

In general, we see that the most prominent type of instability in our simulations is the ejection of a planet  (2.5\,$\%$; 87/3485) and that occurs mainly in the WD phase. Planet--planet collisions happen at a rate of  0.26\,$\%$ (9/3485), 
Finally,  0.14\,$\%$ (5/3485) of the systems experience a collision between a planet and the star.

In the pre--WD phase  (377.65 -- 477.57  Myr), our simulations resulted in 
4 (0.12\,$\%$) dynamical instabilities,  two planet--planet collisions and two planet ejections. It is worth to mention that we have obtained 15 simulations where both planets are stable on the MS but they have orbit crossing just at the AGB tip of the host star, changing the order of the planets at the beginning of the WD phase.

During the WD phase  97 (2.78\,$\%$) of the simulations resulted in the loss of a planet and there most of them,  85 (2.44\,$\%$) correspond to ejections, a few 7 (0.2\,$\%$) are collisions between the planets and just 5 simulations (0.14\,$\%$) resulted in a direct collision of the planet with the WD. 

To end this sub--section, it is worth mentioning that \citet{veras2013b} performed a test as to whether two--planet systems discovered up to November 2012 remain 
stable using their Hill stability criterion and assuming planets with minimum mass and coplanarity. 
They found four unstable planet pairs of two--planet systems, namely,
24~sex, HD~128311, HD~200964,   which we also found  to be unstable (during the MS phase) because they require stabilisation by 2:1 and 4:3 mean motion resonances; the fourth system BD+20~2457 is not analyzed in this work since it is an evolved star. Additionally, they analyzed three more pairs of planets which are expected to be Hill stable but Lagrange unstable in late MS times, HD~183263, HD~108874 and HD~4732. In our 10 Gyr simulations,  HD~183263 indeed becomes Lagrange unstable at MS but using the updated parameters we find it stable. HD~108874 remained stable the entire simulated time and HD~4732 was not considered in this work since it is a giant star.

\subsection{The planet semimajor axis ratio}
\label{semitime}

\begin{figure}
\begin{center}
\includegraphics[width=9.3cm, height=8cm]{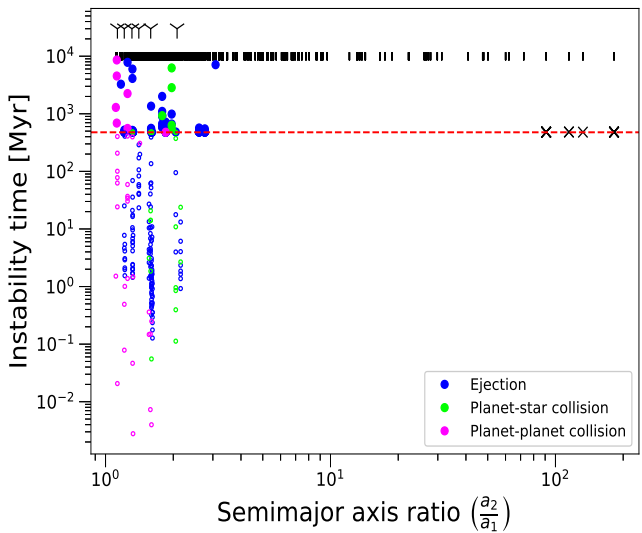} 
\caption{Instability time as a function of semimajor axis ratio of our two--planet systems.  Ejections are shown in dark blue, planet--planet collisions in pink and planet--star collisions in light green. The black vertical ticks mark the semimajor axis values of the two--planet systems used in this study. First-- and second--order mean motion commensurabilities (6:5, 4:3, 3:2, 5:3, 2:1, 3:1 from left to right) are shown with Y--shaped black symbols at the top of the graph. The red horizontal dashed line marks the time where the star becomes a WD. Instabilities in the MS are shown with small empty circles. The x--shape symbols reflect planet ejections produced when the non--adiabatic mass loss regime is reached. }
\label{instabili}
\end{center} 
\end{figure}

In the following we perform a deeper analysis of the two--planet system parameters and see how they relate with the instability times obtained in our simulations. We begin by analyzing the semimajor axis ratio. In Figure \ref{instabili} we show the instability time vs the semimajor axis ratio of our two--planet systems, together with location of some first-- and second--order mean motion commensurabilities. The instabilities on the MS occur near these commensurabilities, most being around the 2:1 confirming the reduced survival rate found in \citet{pu2015} with {\it Kepler} multiple--planet systems due to having planets in first and second order commensurabilities. On the other hand, \citet{veras2013b}  found that their simulations of two--planet systems, using planets with identical masses and eccentricities $e_1=e_2 = 0.1$, become completely stable inside the 2:1, $a_2/a_1 = 1.58$ for a $3\mathrm{\,M}_\odot$ host star. In our simulations, we find that the most widely--separated systems that lose a planet on the MS lie near to the 3:1 mean motion commensurability, at a semimajor axis ratio $a_2/a_1 \approx 2$.

We find four planetary system (WASP--53, HD~187123, HD~219828 and PR0211)  for which 17 simulations resulted in the ejection of planet--2 just within the first 6 Myr after the beginning of the WD phase. The main characteristic of these systems is that they have the largest semimajor axis ratio, then, due to our scaling set--up, their planet--2 is initially located at 909, 1147, 1324 and 1821 \,au, with eccentricities of 0.84, 0.252, 0.812, and 0.7 respectively. \citet{veras2011} have found that one planet located at distances of $\sim$ 1000 \,au, having such large, eccentric orbits, may enter the non--adiabatic  mass loss regime of the host star (which means that the mass loss time--scale is comparable to the planetary orbital time--scale); then, the wide and eccentric planet enters to a run--away phase, where its eccentricity and semimajor axis increase drastically, resulting in the ejection of the planet at times when the star has lost at least 70\,$\%$ of its mass (the fraction of mass lost by a 3$M\odot$ star when it becomes a WD). Alternatively, it can be protected from ejection, depending on its true anomaly evolution. With this in mind, the planet ejections found in the four scaled system mentioned before are due to the non--adiabatic mass loss and not by Lagrange instability. We show these planet ejections in Figure \ref{instabili} with x--shaped symbols.

The theoretical  Hill stability limit in the WD phase in terms of semimajor axis ratio is calculated using the procedure found in \citet{veras2013b} and can be compared to the  parameters of the simulated systems. We calculate the ratio between the observed semimajor axis ratio and the WD Hill stability limit. From our  3485 simulations of two planets, we found that  106 have an observed semimajor axis ratio lower than the expected ratio at which they might be Hill--stable at the WD phase. We can expect these systems to be unstable in the MS or  WD phases. After performing the simulations, we found that 51 out of the  106 of those systems have dynamical instabilities where a planet is lost by ejections, planet--planet collisions or planet--star collisions, in the pre--WD or WD phases.  17 simulations undergo orbital scattering in their planet orbits and/or orbit crossing between the planets, without losing a planet. The other  38 simulations, of the  106 predicted unstable, remained stable the entire simulated time.  Nevertheless, regarding the simulations expected to be stable the entire simulated time, we obtained 33 simulations which were Hill--stable during the WD phase but Lagrange--unstable, losing a planet in this phase. The latter cases are clear examples of planetary systems Hill--stable that become Lagrange--unstable due to the mass loss of the star, confirming that the boundaries of stability changes as the star becomes a WD \citep{debes2002}. We also have  15 simulations expected to be Hill--stable showing orbital scattering in their planet orbits during the WD phase without any orbit crossing nor loss of a planet. Note that we are not counting amongst the latter the 17 simulations with planet ejections produced by the non--adiabatic mass loss regime. 

\begin{figure}
\begin{center}
\includegraphics[width=9cm, height=7cm]{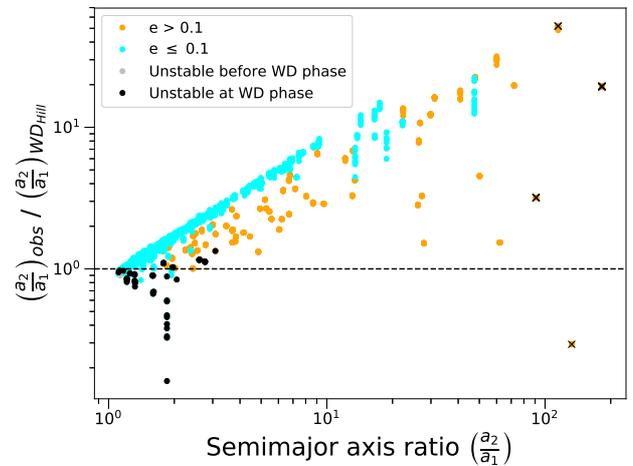} 
\caption{Ratio of the simulated semimajor axis ratio of the two--planet systems and the theoretical Hill limit calculated following the prescription in \citet{veras2013b} for the WD mass, as a function of the simulated semimajor axis ratio. Black dashed line depicts where the simulated semimajor axis ratio and the semimajor axis ratio of the Hill limit at the WD phase are equal.  In light blue dots we show the systems for which both planets have eccentricity $\leq$ 0.1. In orange dots planets with one or both planets have eccentricities $>$ 0.1. Gray dots marked the unstable systems in the simulations and black dots show when the instabilities happen at the WD phase. The x-shape symbols depict simulations where ejections of the planets so far away that enter into the non--adiabatic mass loss regime.}
\label{mudelarat}
\end{center} 
\end{figure}

In Figure \ref{mudelarat} we show the ratio between the observed semimajor axis ratio of the planetary systems and the critical semimajor axis limit at which the planets may become Hill unstable, using the theoretical prescription given by \citet{veras2013b} with the WD mass, plotted as a function of observed semimajor axis ratio. 
We observe that most of the two--planet systems follow a trend that indicates the larger the semimajor axis ratio, the larger the difference of the observed ratio with respect to the theoretical one. We note that the light blue dots, representing low--eccentricity systems, follow a straight line with positive slope. Nevertheless, for planets that have eccentricities higher than 0.1, their semimajor axis difference deviates from this trend.

\subsection{The planet:star mass ratio}

We analyze effects of the the planet:star mass ratio defined as $\mu=\frac{m_1+m_2}{M_*}$, where $m_1$ and $m_2$ are the planet masses and $M_*$ is the mass of the central star. We use this to calculate  the separation of the planets in mutual Hill radius units defined as,
\begin{equation}
R_\mathrm{H,mutual}=\frac{a_1+a_2}{2}\left(\frac{m_1+m_2}{3M_*}\right)^{1/3}    
\end{equation}
where $a_1$ and $a_2$ are as stated before the semimajor axis of planets 1 and 2 respectively. Note that here we use the mass of the host star as $3\mathrm{\,M}_\odot$ for the calculation of the mutual Hill radii.

\begin{figure}
\begin{center}
\includegraphics[width=9cm, height=7.5cm]{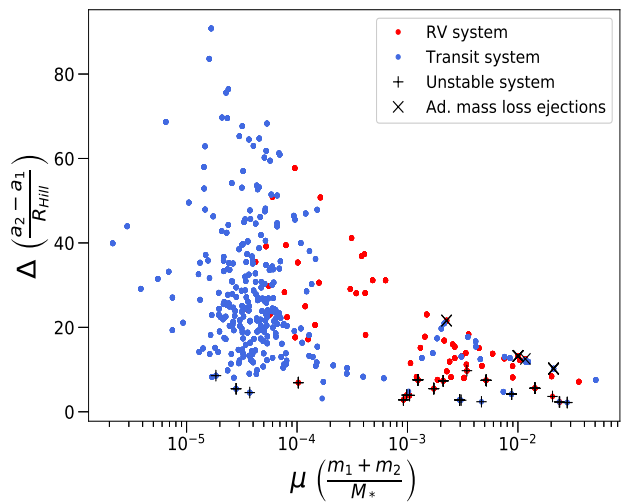} 
\caption{Distribution of planet separation distance in Hill radii ($\Delta$) as a function of $\mu$ $((m_1+m_2)/M_*)$, where $M_*$ is the mass of the host star in the MS. Red points mark systems discovered by RV, blue those by transit. Systems where at least one run was unstable during the WD phase are represented as black plus symbols, and those experiencing ejections by non--adiabatic mass loss as x--shape symbols. }
\label{mudel}
\end{center} 
\end{figure}

Since the distribution of separations of the planets $\Delta$ in terms of mutual Hill radius is also a function of $\mu$, in Figure \ref{mudel} we show $\Delta$ as a function of $\mu$ for the two--planet systems analyzed in this work.

In general, we see that lower--mass planets can be more widely spaced in mutual Hill radii. The envelope in the $\mu-\Delta$ plane is simply related to the fact that there is a singularity in the relation between $\Delta$ and $\mu$, where $\Delta_{\mathrm{max}}=2(\frac{m_1+m_2}{3M_*})^{-1/3}$ \citep[cf.][]{mustill2014}. For a better understanding of the planet masses in terms of $\mu$, planets with Earth, Neptune and Jupiter masses have values of the order of $10^{-6}$,  $10^{-5}$, and  $10^{-3}$ respectively. We highlight that the unstable systems that lose a planet by Hill or Lagrange instabilities are located in the lower part of the graph ($\Delta\leq$   9.74) 
and in the $\mu$ range from $1.8\times10^{-5}$ to $0.03$.

\begin{figure*}
\begin{center}
\includegraphics[width=17.5cm, height=7.7cm]{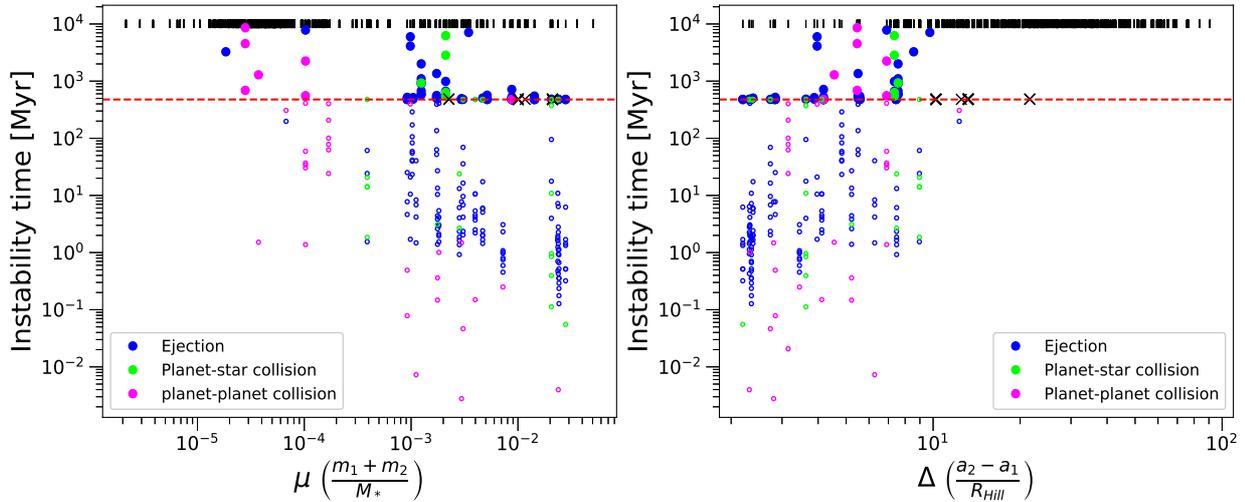} 
\caption{Left: Instability times vs $\mu$. Colors and symbols are as in Fig. \ref{instabili}. Right: Same as left panel but showing the separation $\Delta$ in terms of Hill mutual radius.}
\label{hilmut}
\end{center} 
\end{figure*}

It is clear that the planetary systems that have been discovered mainly by the transit method differ in their distribution in the $\mu$-$\Delta$ space with respect to those detected by RV (see Figure \ref{mudel}). 
The majority of the planets detected by transits are less massive than those detected by RV, and therefore they have smaller Hill radii. This means that they can be very widely dynamically spaced (in terms of Hill radii) even when they are rather closely spaced physically (in terms of \,au or semimajor axis ratio). This means that it is easier for the transiting systems to remain stable than it is for the RV systems. We demonstrate this statement in the left panel of Figure \ref{hilmut}, where we plot the instability time (Myr) vs $\mu$. Most of the instabilities occurring during the MS phase are due to planets with masses higher than that of Jupiter. In fact, the instabilities happening at very early times (between 0.1 to 10 Myr) are those produced for the high mass planets ($m_\mathrm{planet} > 1 \mathrm{M_J}$). Then, as $\mu$ decreases, the instabilities move to later times, and at some point ($\mu\leq 1.84\times10^{-5}$), the instabilities cease to happen. In the right panel of Figure \ref{hilmut} we see how most of the MS and WD instabilities that lead a loss of a planet happen at $\Delta \leq$ 10 and then the planets are separated enough so that they were stable the entire simulated time.

\subsection{The planet mass and eccentricity ratio}

\begin{figure*}
\begin{center}
\includegraphics[width=17.5cm, height=7.7cm]{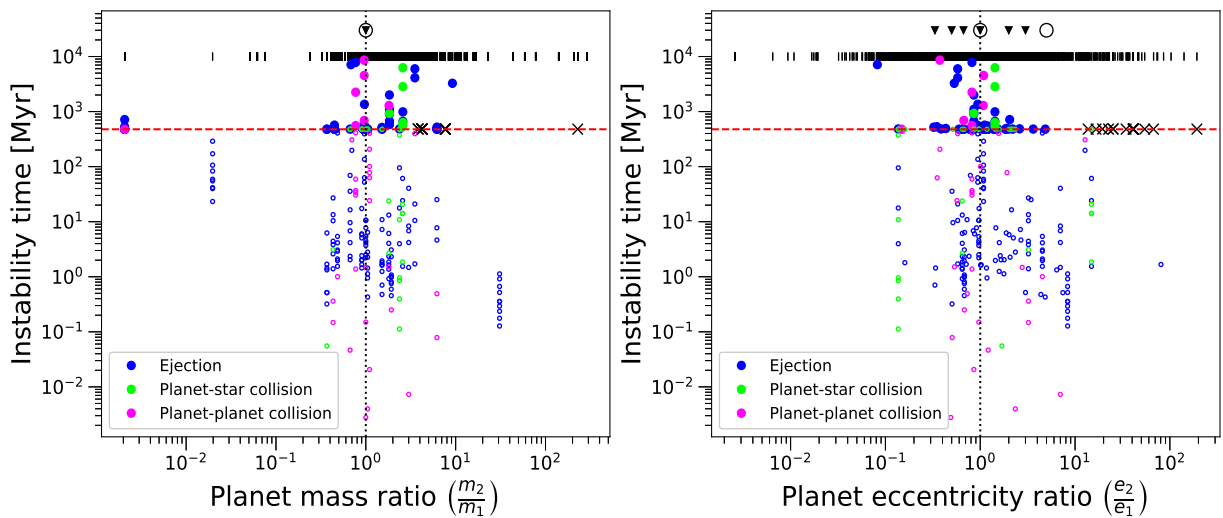}
\caption{Left panel. Instability times vs the planet mass ratio. Colors and symbols are as in Fig. \ref{instabili}. Right panel. Instability times as a function of the eccentricity ratio. Black circles and upside down triangles in the top of both panels mark the planet mass ratio (right) and eccentricity ratios (left) of two--planets simulations used in  \citet{veras2013b,veras2013} respectively. The black dotted line depicts the ratios for which the mass and eccentricity is equal in both planets.}
\label{massecc}
\end{center} 
\end{figure*}

In the following we explore how  the mass and eccentricity of the planets relate to the instability times. In the left panel of Figure \ref{massecc}, we plot the instability times (Myr) vs the planet mass ratio. We have also shown the cases explored in the simulations performed by \citet{veras2013b,veras2013}. 

We can see that most of the  systems that have an instability at the WD phase are located in the range between 0 to 9.2 planet mass ratio. The most extreme is Kepler--487, which has a planet mass ratio of $\sim 2\times 10^{-3}$. For this system, the mass--radius calculation of planet--1 for a 10.9 $R\oplus$ gives a planet mass of 2595 $\mathrm{M_\oplus} ($8.16 $\mathrm{M_J}$), while, for planet--2 the calculated mass is 5.25 $\mathrm{M_\oplus}$ for a radius of 2.07 $R\oplus$. We clearly see that in most of our simulations, planet--2 is more massive than planet--1, verified by having a large number of black lines in the right of the dotted line (and see Figure~\ref{mrati}). The number of instabilities at the WD phase for a planet mass ratio $<$ 1 is  slightly larger than those for a mass ratio $\geq$ 1 (44 and  36 respectively).

In the right panel of  Figure \ref{massecc} we show the same as in the left panel but as a function of the eccentricity ratio, where black upside down triangles depict the eccentricity ratios studied in \citet{veras2013}. The instabilities leading a loss of a planet at the WD phase in our simulations are present in a wider range of eccentricity ratios from the ones explored before in the literature, from 0 up to  4.86. We note that in contrast to the planet mass ratio, the eccentricity ratio values explored in our simulations are quite symmetric with respect to the ratio where the eccentricity of both planets is the same. We also note that the planet ejections due to non--adiabatic mass loss are located with eccentricity ratios between 10 to 100, even one case with a value around 200, which means that those systems have planetary configurations of a very eccentric planet--2 with respect to its companion. Nevertheless, we find more planet losses at the WD phase at eccentricity ratios $\geq$ 1  than $<$ 1 (43 simulations vs 37).        

\section{DISCUSSION}

\label{planetesi}

\subsection{Instability without loss of a planet}

Hitherto we have treated systems as ``unstable'' when they lose a planet due to ejection or collision. However, systems where planets experience some degree of scattering without being lost are also of relevance for polluting white dwarfs, as their changing orbits and increasing eccentricities can lead to scattering of asteroids. Indeed, \cite{mustill2018} found that such systems can be among the most efficient at delivering material to the WD.

Among these systems, there are three groups to identify. The first group consists of  26 simulations in which orbits intersect and this produces orbital scattering in the semimajor axis until one of the planets is lost, either by ejection, planet--star collision or planet--planet collision at the WD phase (13 of them have orbits crossing  before the WD phase). In the second group we include the  14 simulations where orbit crossing and orbital scattering in the semimajor axis are present, but no planets are lost in the 10 Gyr of simulated time (2 simulations have orbit crossing  before the WD phase). In the third group of  18 simulations the planet orbit do not cross and no planet is lost either,  but they show orbital scattering in the semimajor axis. The orbital scattering is defined as those systems where the observed semimajor axis of planet--1  and/or planet--2 differs more than   5\,$\%$ with respect to the semimajor axis value at the beginning of the WD phase. In total we find that  58 simulations out of  3485 (1.66\,$\%$) are located within the groups defined before. Note that in the 54 simulations out of  80 where a planet is lost by Hill or Lagrange instabilities during the WD stage do not experience any previous orbit crossing.

\begin{figure*}
\begin{center}
\includegraphics[width=8.5cm, height=6.9cm]{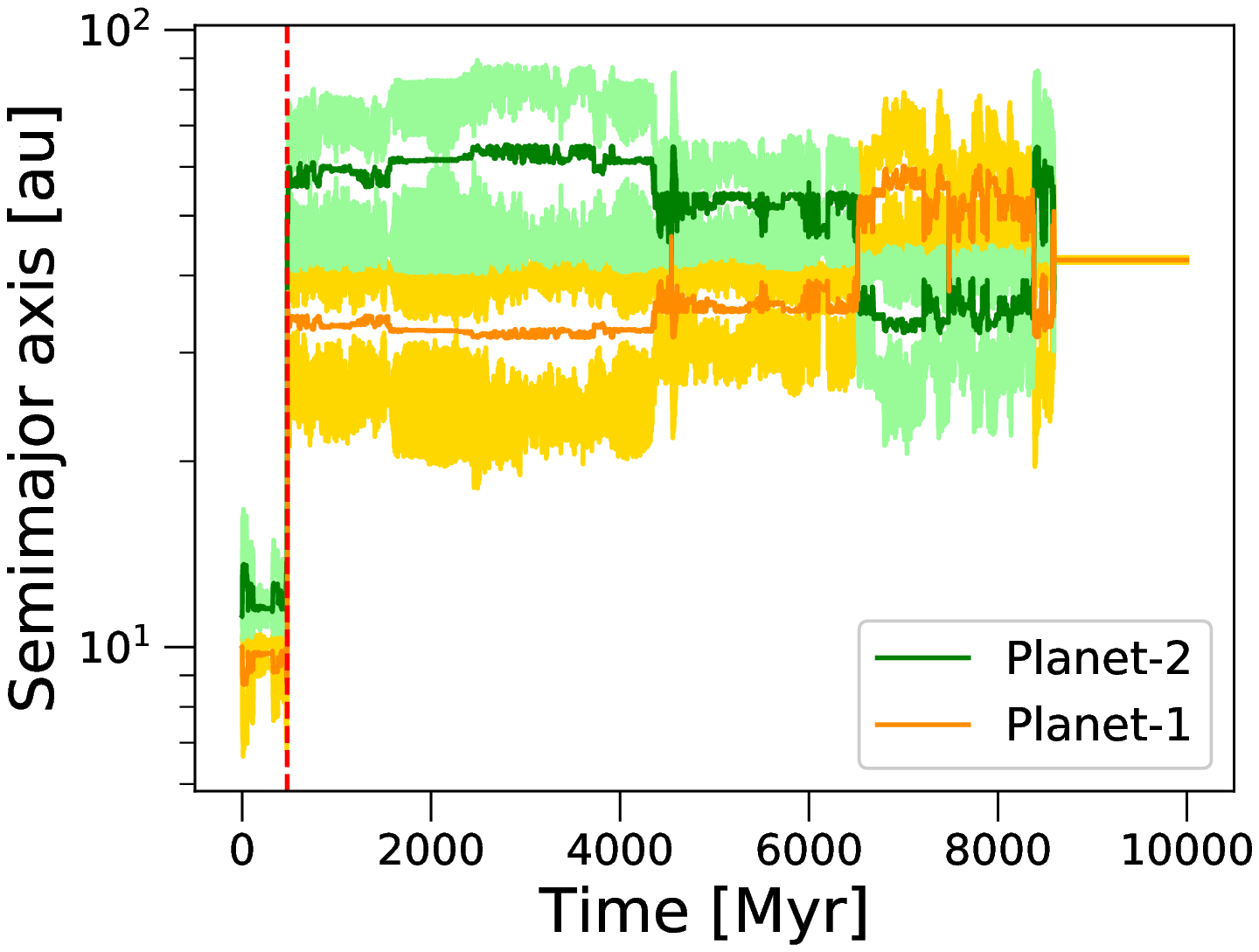}
\includegraphics[width=8.5cm, height=6.9cm]{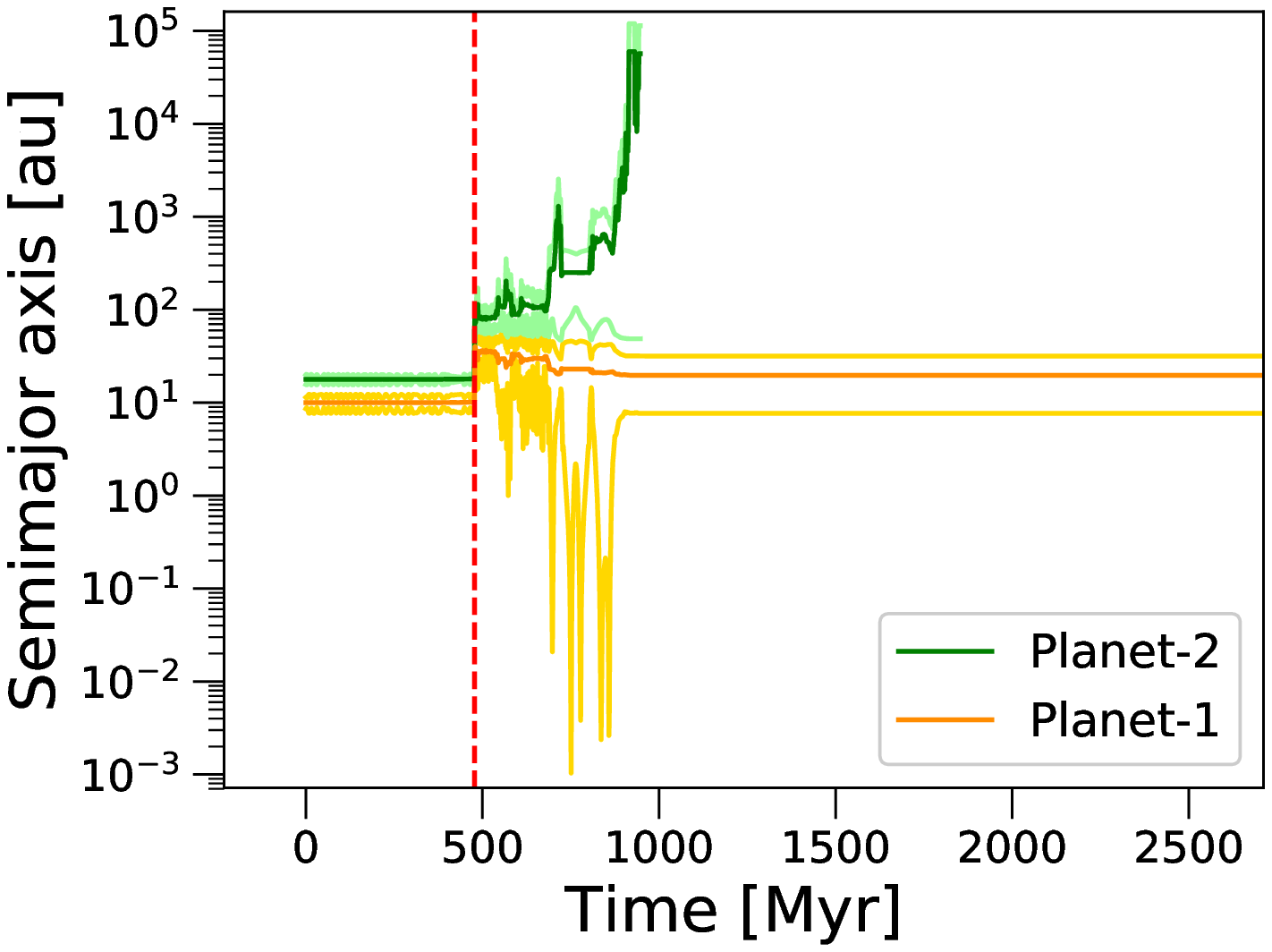}\\
\includegraphics[width=8.5cm, height=6.9cm]{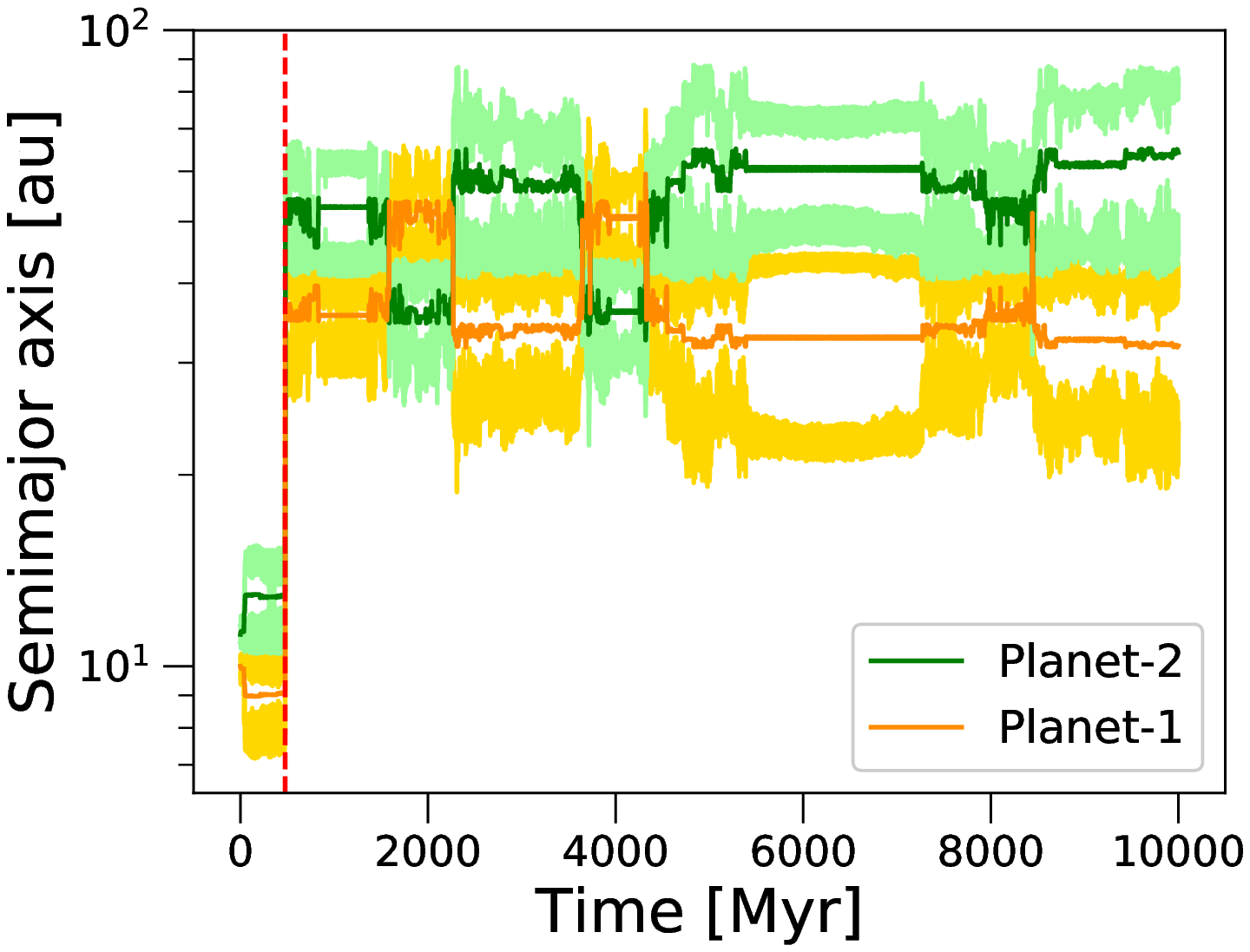}
\includegraphics[width=8.5cm, height=6.9cm]{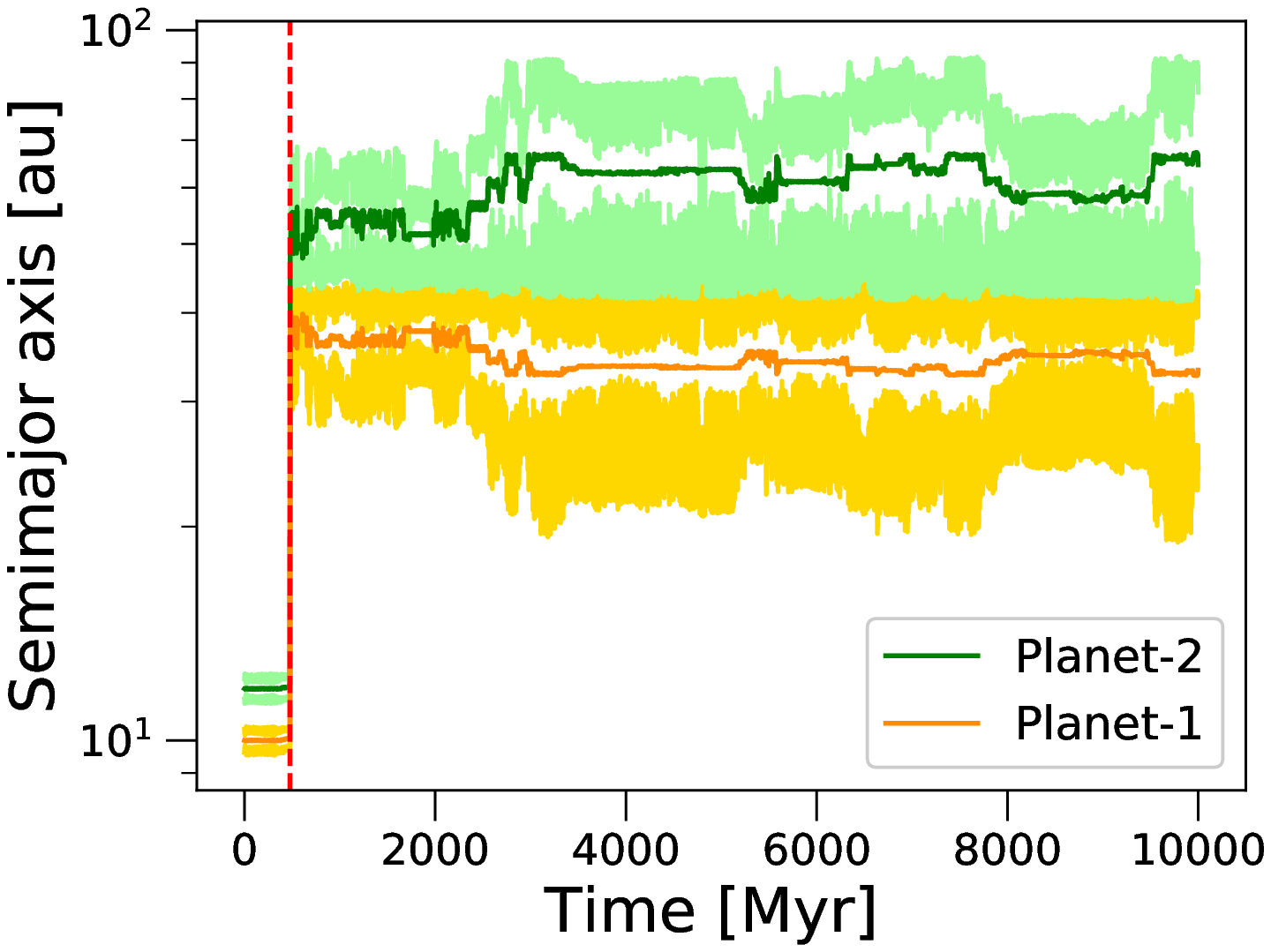}

\caption{Four representative examples of orbital evolution where the  orange and  green lines show the semimajor axis of planet--1 and planet--2 respectively. The lighter versions of these lines show the evolution of the apocenter and pericenter of the planets. As usual the red dashed vertical line represents the time when the star becomes a WD.   From top to bottom and left to right, the input parameters of these simulations can be found in Table \ref{mms} under \#1831, 866, 1836, and 2352. }
\label{scat}
\end{center} 
\end{figure*}

In Figure \ref{scat} we display four examples of the groups defined previously, where the evolution of the semimajor axis is shown as a function of time. We have added in the plot the evolution of the apocenter and pericenter. In the upper panels we show the orbital evolution of the scaled systems {\it Kepler}--200 (left) and HIP~65407 (right), both of them losing a planet by planet--planet collision and ejection respectively; {\it Kepler}--200 also has the orbits of its planets crossing several times before the planet--planet collision. We highlight for example that the top left simulation exhibits a collision between the planets at 8.6 Gyr, which would produce debris that can be launched toward the WD by the remaining planet.  The system of the top right experience the ejection of planet--2 at 0.95 Gyr, however,  we highlight that the pericenter of the surviving planet--1  reaches the Roche radius of the WD several times, the first time at 751 Myr and the last time at 859 Myr (see section~\ref{sec:roche} for further discussion). In the lower panels we present another run of the scaled systems {\it Kepler}--200 (right) and the system {\it Kepler}--29 (left). In these cases both planets are bound to the system the entire simulated time, however in {\it Kepler}--200 there are several instances of orbit crossing during the 10 Gyr, while {\it Kepler}--29 does not have any orbit crossing but the orbital scattering is quite large, increasing as the system evolves up to distances where the pericenter and apocenter of the planets cover a range of $\sim$ 80 \,au for the last Gyr of the simulation. Note that the scattering in the simulation in the bottom left panel begins during the MS and following the mass loss this orbital behavior causes the planets to migrate from the inner to the outer regions of the planetary system. While the case shown in the bottom right there is not a planet loss nor an orbit crossing between the planets but there is a quite large orbital scattering in the pericenter and apocenter of planet--1 and --2 for the entire 10 Gyr of simulation. 

\subsection{Reaching the Roche limit of the WD}

\label{sec:roche}

The simulation of the scaled system HIP~65407 shown in the upper right panel of Figure \ref{scat} serve as an example to highlight a very important behaviour in the simulations: clear scattering is present, especially in the pericenter of planet--1, but most importantly there are  periods of time when the pericenter of planet--1 reaches very close distances to the WD radius (0.02 $R_\odot$). During these instances, the planet could experience tidal destruction, an effect not included in the simulations. Therefore, we post--process the simulation results in order to compare the pericenters of all the planets with the Roche radius of the WD,
\begin{equation}
a_\mathrm{Roche}=\left(\frac{3\rho_\mathrm{WD}}{\rho_\mathrm{pl}}\right)^{1/3}R_\mathrm{WD}    
\end{equation}
where $\rho_\mathrm{WD}$, $\rho_\mathrm{pl}$ are the densities of the WD and the planet respectively and $R_\mathrm{WD}$ is the radius of the WD \citep{mustill2014}. 

 Seven of our simulations have the pericenter of planet--1 at smaller distances than the calculated Roche radius:  two of the system HIP~65407 (one of them shown in the upper right panel of Figure \ref{scat}), and  five in the system HD~113538. These occur at cooling ages of $\sim100$\,Myr to several Gyr.  The five planet--star collisions found in the WD phase are within the 7 simulations where planet--1 crosses the Roche radius. This means that these planets could be tidally disrupted by the WD, hence producing a circumstellar disc and pollute its atmosphere through acreted material.

\citet{gansicke2019} recently interpreted the gas disc in WD~J0914+1914 as the photo--evaporated atmosphere of an icy giant planet. They reached this conclusion from the inconsistency of accretion from the wind of a low mass stellar companion to WD~J0914+1914, the depletion of rock--forming elements with respect to bulk Earth composition, and the larger size of the circumstellar disc compared with a canonical disc that a planetesimal would form. Also, they also argued that Neptune to Jupiter--mass planets need to be at distances lower than 14--16 $R_\odot$ so that the planet atmospheres begin to photo--evaporate and form a gas disc similar to the WD~J0914 disc. With this less restrictive condition in mind, we check whether we find more systems with pericenters reaching this $16R_\odot$ limit. We find  2 more simulations:  one of the scaled system HD~113538, where one planet is ejected at 988 Myr and one of the system HD~30177, where the planet is lost by ejection at 547 Myr. Note however that for evaporation to take place high irradiation during extended periods of time is a requirement (see e.g. \citealt{Villaver2007}) and that condition is hardly full filled just by reaching the pericenter distance mentioned above. Nevertheless, some of these close pericentre passages in our simulations occur at cooling ages of just a few 10s of Myr, when the WD is still indeed hot and bright.

\begin{figure*}
\begin{center}
\begin{tabular}{cc}
\includegraphics[width=9cm, height=7.5cm]{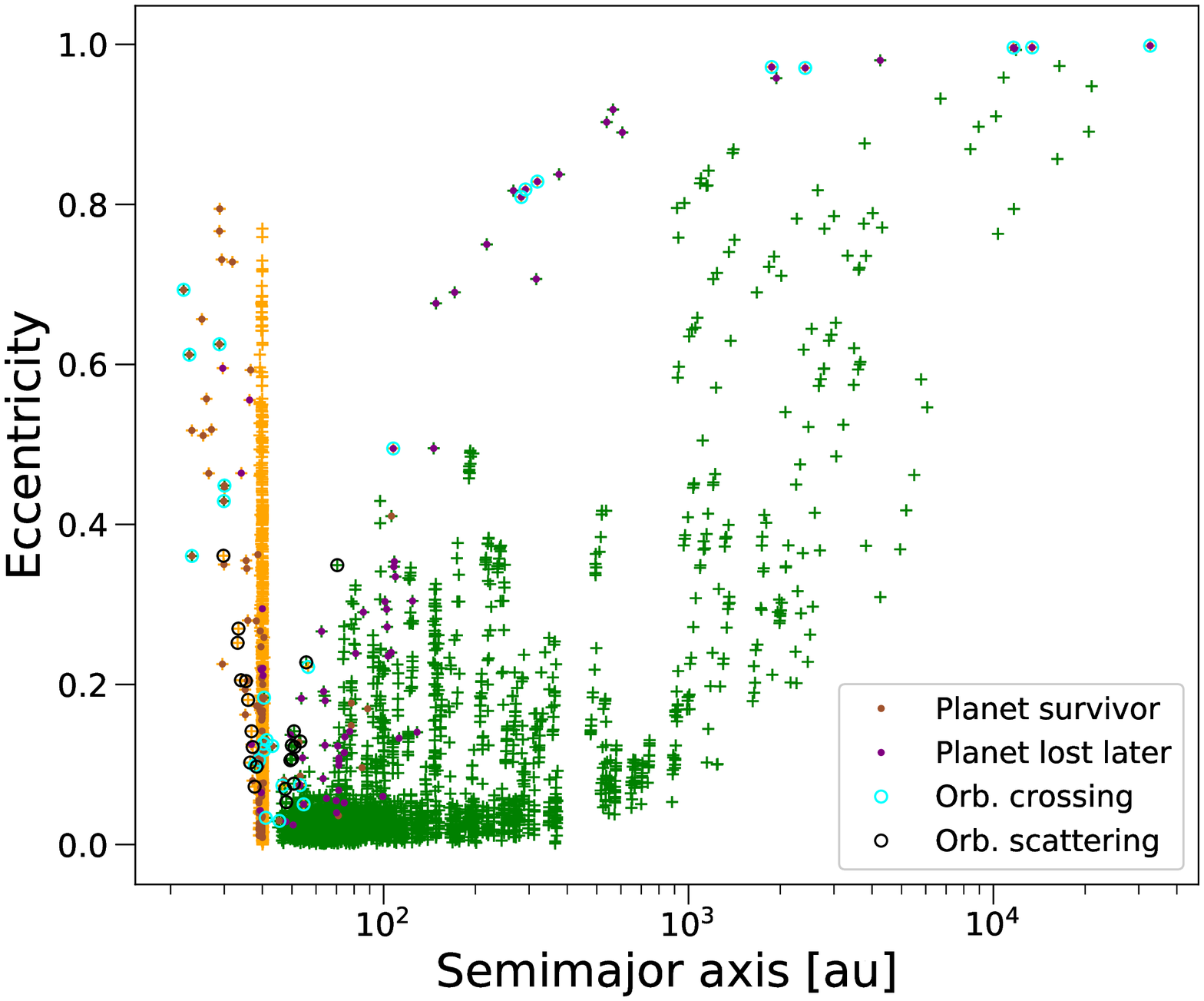} 
\includegraphics[width=9cm, height=7.5cm]{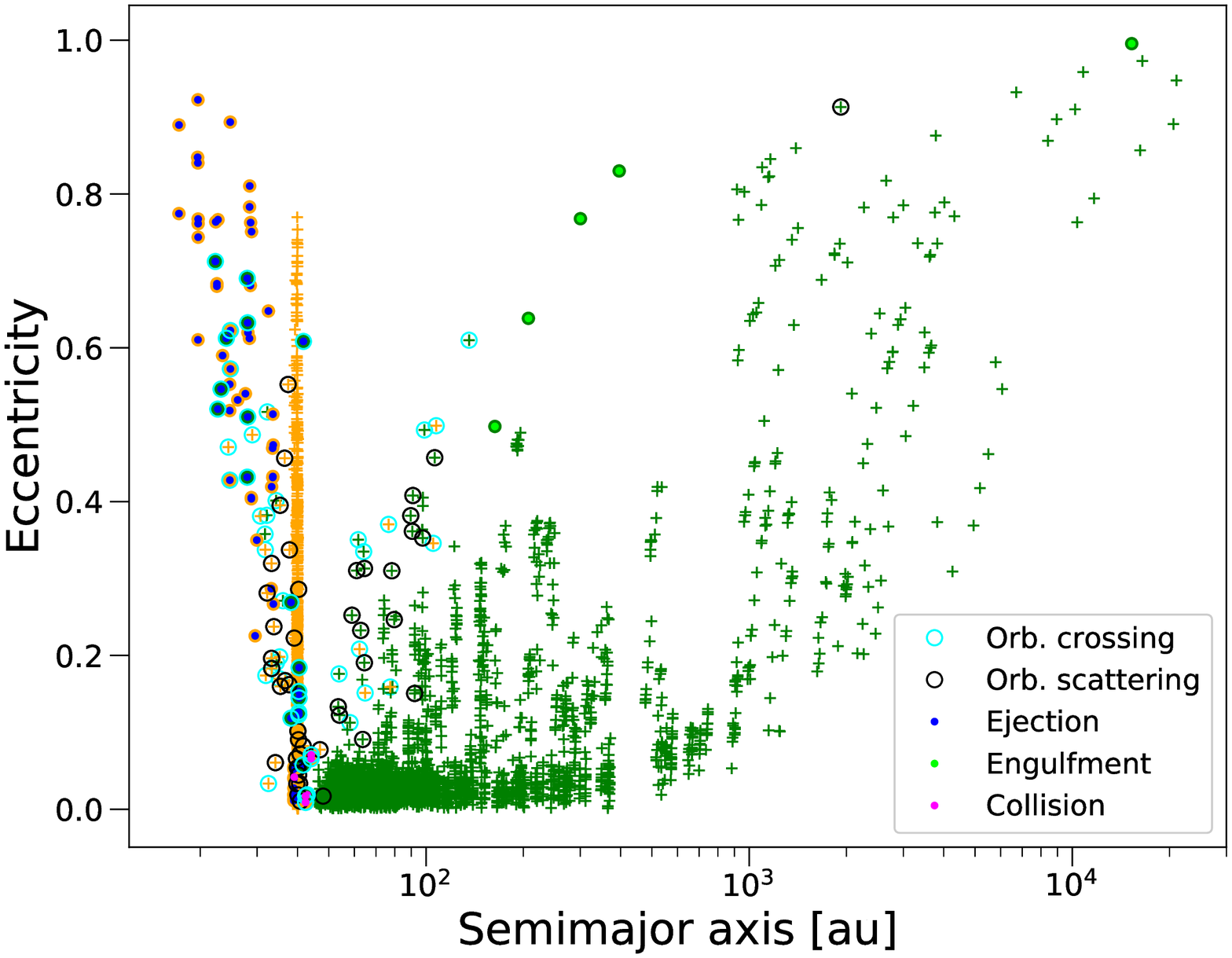} 
\end{tabular}
\caption{Eccentricities vs semimajor axis of the scaled two--planet systems simulated in this work in the left panel at the time immediately after the central star becomes a WD (478 Myr) and the the right 
at the end of the simulations (10 Gyr). The  orange and  green colours are for planet--1 and planet--2 respectively and the plus symbols are for planets in systems that never lose a planet. Brown dots indicate  planets that survive the loss of their companion, while the purple ones depict those planets that will be lost during the WD times. Light blue and black circles around the plus symbols depict planets that undergo an orbit crossing before the WD time and planets that have orbital scattering without any loss of a planet nor orbit crossing respectively. The 10 planets that are unstable just or around that time (given the time resolution of the outputs of the simulations) are not shown in the graph nor the 17 planets ejected due to non--adiabatic mass loss.  Right panel:  we highlight the planets that lost a planet companion after the system suffered a dynamical instability in the WD phase. In dark blue dots we show the planets that survived the ejection of the other planet, in pink dots the planets surviving a planet--planet collision, and in light green dot planets surviving a planet--star collision of their companion. The light blue and black circles show the same as the left panel with the difference that now indicate orbit crossing or orbital scattering only at the WD phase.}
\label{surveccar}
\end{center} 
\end{figure*}

\subsection{Eccentricity and Planet mass}

In the two snapshots of Figure \ref{surveccar} we show the distribution of the final eccentricity vs final semimajor axis of our two--planet simulations: on the left at the beginning of the WD phase (478 Myr), and on the right at the end of the simulations (10 Gyr).  
This is to be compared with the initial $a-e$ distribution shown in Figure~\ref{initea}.

Planet--1 is mostly found at $\sim40$\,au, as expected from adiabatic orbit expansion, with a secondary tail of planets that have experienced scattering extending in to 20\,au at a range of eccentricities. This tail is typical for orbital elements after scattering \citep[e.g.,][]{chatterjee2008,mustill2014}.
The only few planets--1 that are beyond 50\,au, when the WD begins are those that underwent orbit crossing at  pre-WD times. Planet--2 fills a much wider region of the eccentricity -- semimajor axis space with planets reaching up distances $\geq$ 10\,000\,au and eccentricities $\geq 0.95$. Note that many of these are the planets that will be ejected later on. The locations of stable planets in Figure \ref{surveccar} are as expected given adiabatic mass loss and neglecting tidal forces \citep{Villaver2007}, $a_\mathrm{WD}=a_\mathrm{MS} (M_\mathrm{MS}/M_\mathrm{WD})$, where $a_\mathrm{MS}$ is the initial orbital radius and $M_ \mathrm{MS}$, $M_\mathrm{WD}$ are the masses of the star at the MS and WD phases respectively. Thus given the initial configuration stable planets--1 are expected to end at 40\,au after mass loss with planets--2 further out, having a bulk of them located at distances from 45 to 350\,au and others beyond $10^3$\,au.
Planets--2 that have so far survived scattering are found in a second tail that extends to high semimajor axis and eccentricity. Planets are not found in the middle between the tails, since they are prone to experience dynamical instabilities in this region \citep{chatterjee2008}.

In the right panel of Figure \ref{surveccar} at 10 Gyr we have added the information of the surviving planets: specifically, what type of dynamical instability leads to the loss of their companions at the WD phase.
We can see that all the surviving planets that lose their companions by ejection ended between 17 to 41\,au, and between $\sim 0$ to $0.92$ in eccentricity, while those surviving a planet--planet collision ended between 39 to 44\,au, with eccentricities $\leq$  0.07 and the survivors of a planet--star collision are wide planets with semimajor axis $> 150$\,au and eccentricities $> 0.49$.

\begin{figure}
\begin{center}
\includegraphics[width=9cm, height=6.5cm]{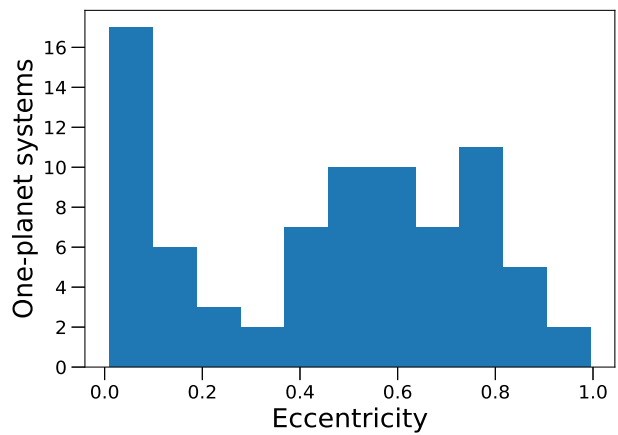} 
\caption{Distribution of eccentricities of the surviving planets after the other become unstable and is lost from the system. The eccentricities are taken once the remaining planet becomes stable at the WD phase.}
\label{eccentric}
\end{center} 
\end{figure}

The distribution of eccentricities of the surviving planet of the initial pair the end of our simulations is shown in Figure \ref{eccentric}.  
We see a bi--modal distribution with one peak at eccentricities around $0.0$ and a second one around $0.6$.
Note that this distribution in the second peak is very different from the initial one (see  Figure \ref{esemi}) meaning that high eccentric planets are produced by interactions of multiple planets. This is important given that simulations that include the interaction of planets with a planetesimal belt conclude that highly eccentric planets may deliver efficiently material toward the WD \citep[see, e.g.,][]{bonsor2011, frewen2014,mustill2018}. Here we provide a mechanism for a planet to have large enough eccentricities to be an efficient deliverer of material: a planet--planet scattering in a multiple planetary system.

We have  80 simulations in which a planet is lost  either by Hill or Lagrange instability during the WD phase. We now look at the distribution of eccentricities of the remaining planet and compare them with the results of previous works. 
We find that in  52 out of this  80 (1.49\,$\%$ of the total  3485) simulations, one of the planets ends up with eccentricities larger than $0.4$. These could be ideal candidates for sending large numbers of asteroids towards the WD; however,  51 of these  52 planets have masses $>1\mathrm{\,M_J}$ and would therefore have low efficiency at delivering material to the WD, preferentially ejecting asteroids instead.

However, we do find lower--mass planets if we include systems which, during the WD phase, experience either i) the loss of a planet, ii) orbit crossing without the loss of a planet, or iii) orbital scattering in the semimajor axis of both planets. 
 112 out of  3485 simulations fulfill this condition (3.21\,$\%$). 
If we now look into their mass distribution we have  25 simulations (0.72\,$\%$) where planet--1 masses are in the range $1-30\mathrm{\,M}_\oplus$ and  35 (1.004\,$\%$) where planet--2 has masses in the same range.
These numbers are  39 (1.36\,$\%$) and 49 (1.41\,$\%$) of simulations for planet--1 and --2 respectively, with masses between 10 to $100\mathrm{\,M}_\oplus$. These mass ranges, especially the lower Earth--Neptune mass end, are those identified by \cite{mustill2018} as being most efficient at delivering asteroids to the WD during and after an instability.

\subsection{Pollution and cooling times}

We now briefly discuss the cooling times at which our 
instabilities occur, and relate this to observations. 
Observationally, polluted WD with detected IR excesses may
have a peak distribution in cooling ages around 400--500 Myr 
while the polluted WDs without IR excesses have a distribution
in cooling times that extends up to 1 Gyr\footnote{Using data
from \citep{farihi2009,debes2011,girven2012,rocchetto2015}
complemented by the WD Montreal Database \citep{dufour2017b}
http://www.montrealwhitedwarfdatabase.org.}. While we do not
perform a quantitative comparison with our simulations, which
would require addressing the biases of different surveys, 
these data do indicate that planetary/asteroidal material 
is being delivered to WDs at a large range of cooling ages. 
In particular, regardless of the true time dependence of these phenomena, 
any dynamical delivery mechanism must be capable of providing 
\emph{some} material at late times.

Indeed, we do find that the number of instabilities where a planet is lost in two--planet systems decreases as the cooling time increases, in common with previous dynamical simulations of planetary systems \citep[e.g.,][]{veras2018,mustill2018}.  Compared to \cite{veras2013b}, we have fewer planet--planet collisions (0.14\,\% of instabilities overall compared to their $\sim50\,\%$): we can attribute this to the fact that \cite{veras2013b} simulated perfectly coplanar systems which significantly increases the likelihood of a physical collision when orbits cross. In our simulations most of the instabilities where a planet is lost occur in the first 100 Myr of the cooling time (55), with  14 simulations  losing a planet  between 100 -- 1000 Myr and  11 of them occur at times  $t \geq 1$ Gyr. In contrast, note that \citet{hollands2018} found that the  metal abundances in  WD polluted sample decays exponentially with an e--folding time of 0.95 Gyr, suggesting that the potentially polluting material is being depleted at post MS times.

\begin{figure}
\begin{center}
\begin{tabular}{cc}
\includegraphics[width=9cm, height=8cm]{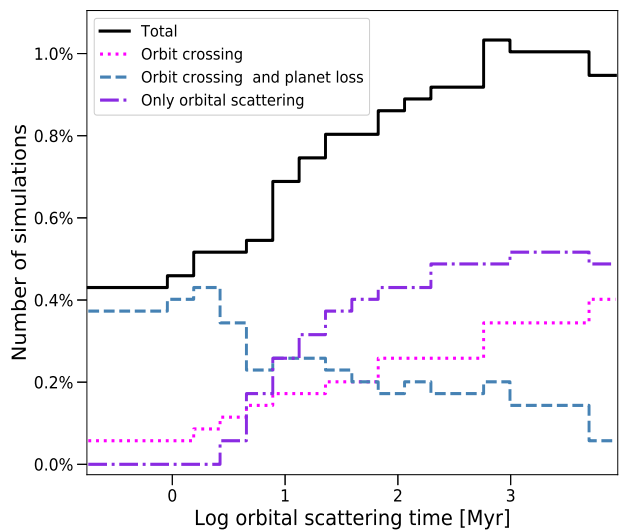} \\
\end{tabular}
\caption{Distribution of WD cooling time of the orbital scattering for simulations in which  orbit crossing appears.  The colours depict different cases. The pink dotted line shows the simulations when the orbit crossing begins and the orbital scattering lasts the total simulated time, since no planets are lost in those simulations. The gray dashed line indicates the simulations from the onset of orbit crossing to the the loss of a planet, either by ejection, planet--star collision or planet--planet collision at WD phase. The purple dash--dotted line display the simulations where only orbital scattering is present, without any orbit crossing nor loss of a planet. The black solid line displays the sum of the pink, gray and purple. We highlight that some systems have the orbit crossing  before the WD phase; thus they start to count from the beginning of the WD cooling time. }
\label{cross_cool}
\end{center} 
\end{figure} 

It is interesting to analyze the two--planet simulations that are dynamically active for periods of time. This period may begin when an orbit crossing is present in the orbital evolution of both planets, since the orbit crossing can trigger long--term orbital scattering in both planets that may launch rocky bodies onto stargrazing orbits, hence producing pollution in the WD atmosphere. This period of time can last until one of the planets in the system is lost, relaxing the system dynamically. However, this interval of time can also last the entire simulated time. 

In Figure \ref{cross_cool}, we display the  percentage of systems currently undergoing orbit crossing and orbital scattering  at different times during the WD phase, there being  58 such systems in total. We transform our simulated time to WD cooling time, setting our zero point at the time when the WD forms. We show separately systems which either do or do not ultimately lose a planet as a result of the orbit crossing and systems where only orbital scattering is present, as well as the sum of these. Some important results can be outlined from this Figure. First, orbit crossing and orbital scattering are typically of long duration, and of the  58 systems at least two thirds are dynamically ``active'', i.e., are between the onset of orbit crossing and the loss (if any) of a planet, at any one time. We see that the peak of simulations where there is a loss of a planet is in the first 10 Myr of cooling time, then it decreases slowly towards the end of the simulation time. This imply that if those systems had an infinite reservoir of planetesimals, the delivery rate would decrease naturally since there would be no events that can launch material toward the WD at Gyr times, then the number of expected Gyr old WD with metal pollution would be very low. On the other hand, by observing the simulations which do not eventually lose a planet, we may expect an increase of number of WD showing pollutant events from hundreds to Gyr times. In reality, we expect that the reservoir of planetesimals is being depleted in time, thus, the effect of observing less WD with pollutant events at older times is enhanced. When we see the sum of the three lines, representing all  58 systems, the predominant effect is  an increase number of pollutant events, reaching a peak around 1 Gyr of cooling time.

\section{Conclusions}

We have explored the stability of multiple planetary systems with the goal of understanding the observed pollution rates detected in WDs. We use dynamical simulations by restricting the, otherwise infinite, parameter space to study the evolution to the WD phase of scaled versions of the MS planetary systems that have been detected. In this way we are exploring a larger parameter space than previous works using configurations of planetary systems with two planets with different masses, orbits, multiple eccentricities and with different semimajor axis ratios. This is the first time dynamical simulations restricted by the observed parameters have been done  to study the instabilities that could bring material to the surface of the WD.   Of course, the only reliable constraint on the planet distribution around massive stars has to come from observations. This will most likely take the form of future microlensing surveys, as might be conducted with the {\it Nancy Grace Roman Space Telescope} \citep{spergel2015,penny2019} that could potentially provide the statistical picture of planetary system architectures on the MS that we need to evolve to the WD to test pollution scenarios. Thus, for the time being we have based our simulations on a simple, well-informed scenario on the observed planet distribution; the scaling of these observations and thus their validity to higher masses does not necessarily reflect reality but is the closest we can explore at the moment with the information available.

We performed 3730 dynamical simulations of 373 planetary systems (we run 10 simulations of each) orbiting a putative 3 $\mathrm{\,M}_\odot$ parent star. After  disregarding for the analysis 245 simulations where a dynamical instability (loss of a planet, orbit crossing) occurs on MS due to having planets in (or close to) mean motion commensurabilities, we ended up with  3485 simulations that we followed for 10 Gyr well into the the WD phase. We find that  80 (2.3\,$\%$) simulations result in the loss of a planet by Lagrange or Hill instability after the formation of the WD, with only 5 of them sending the planet into the WD.  The small number of  planet--star collisions we find confirms that these events are rare in the WD phase, if we consider also the simulations of \citet{veras2013, veras2013b} where they did not encounter any planetary system that resulted in a planet--star collision in the WD phase. It has been estimated that  at most 1--5\,\% of WDs have high ongoing accretion rates due to dust discs \citep{debes2011, farihi2012}. The formation of dust disks around WDs is not followed at all in our simulations but if it is a consequence of instabilities that send the planets themselves into the WD to be tidally disrupted, we find that our two-planet system simulation rates are too low to account for this phenomenon. On the other hand,  if disks are formed after asteroids are scattered and disrupted, following orbital crossing/scattering of the planets, then our overall result (3\,\%) two body dynamics could account for this level of pollution and if it is 100\,\% effective, most or all of the large levels of IR excess due to dust.

We obtain a rate that is so small that implies that other mechanisms have to be invoked to explain the prevalence of atmospheric pollution in at least $25-50\,\%$ WDs. 
One possibility is that higher planet multiplicity plays a role and to explore this option the simulations of the dynamical evolution of observed systems with three and more planets is on--going. As has been shown before, an enhanced incidence of instabilities, increases for systems with three or more planets \citep{mustill2014,veras2016b,mustill2018}. Another important aspect to consider is the fact that planets can be just the mechanism for the delivery of planetesimals and we can increase the delivery rate to   3.21\,$\%$ of our simulations if we assume that orbit crossing and orbital scattering could be contributing to the planetesimal delivery toward the WD. Note that although low, this provides a mechanism to continuously send destabilized material towards the WD.

\section*{Acknowledgements}
This research has made use of the NASA Exoplanet Archive, which is operated by the California Institute of Technology, under contract with the National Aeronautics and Space Administration under the Exoplanet Exploration Program.  This research has made use of the SIMBAD database, operated at CDS, Strasbourg, France.  E.V. and R.M. and A.J.M acknowledge support from the `On the rocks II project' funded by the Spanish Ministerio de Ciencia, Innovaci\'on y Universidades under grant PGC2018-101950-B-I00. MC, RM and EB thank CONACyT for financial support through grant CB-2015-256961. A.J.M. acknowledges support from the the project grant 2014.0017 `IMPACT' from the Knut \& Alice Wallenberg Foundation, and from the starting grant 2017-04945 `A unified picture of white dwarf planetary systems' from the Swedish Research Council. We are grateful to Rafael Gerardo Weisz and Francisco Prada for their critical help with the automation of the processes and the use of the cluster.

\section*{Data availability}

Table \ref{mms} content is fully available as a machine readable table in the online supplementary material.




\bibliographystyle{mnras}
\bibliography{bib} 








\bsp	

\label{lastpage}
\end{document}